\documentclass[iop]{emulateapj}

\usepackage{graphicx}
\usepackage{amsmath}
\usepackage{subfigure}
\usepackage{float}
\usepackage{natbib} 
\usepackage{hyperref}
\usepackage{tabularx}

\begin{document}

\title{EvryFlare II: Rotation periods of the cool flare stars in TESS across half the Southern Sky}

\author{Ward S. Howard\altaffilmark{1}, Hank Corbett\altaffilmark{1}, Nicholas M. Law\altaffilmark{1}, Jeffrey K. Ratzloff\altaffilmark{1}, Nathan Galliher\altaffilmark{1}, Amy Glazier\altaffilmark{1}, Octavi Fors\altaffilmark{1,2}, Daniel del Ser\altaffilmark{1,2}, Joshua Haislip\altaffilmark{1}}
\altaffiltext{1}{Department of Physics and Astronomy, University of North Carolina at Chapel Hill, Chapel Hill, NC 27599-3255, USA}
\altaffiltext{2}{Dept. de F\'{\i}sica Qu\`antica i Astrof\'{\i}sica, Institut de Ci\`encies del Cosmos (ICCUB), Universitat de Barcelona, IEEC-UB, Mart\'{\i} i Franqu\`es 1, E08028 Barcelona, Spain}

\email[$\star$~E-mail:~]{wshoward@unc.edu}

\begin{abstract}
We measure rotation periods and sinusoidal amplitudes in Evryscope light curves for 122 two-minute K5-M4 TESS targets selected for strong flaring. The Evryscope array of telescopes has observed all bright nearby stars in the South, producing two-minute cadence light curves since 2016. Long-term, high-cadence observations of rotating flare stars probe the complex relationship between stellar rotation, starspots, and superflares. We detect periods from 0.3487 to 104 d, and observe amplitudes from 0.008 to 0.216 $g^{\prime}$ mag. We find the Evryscope amplitudes are larger than those in TESS with the effect correlated to stellar mass (p-value=0.01). We compute the Rossby number (R$_o$), and find our sample selected for flaring has twice as many intermediate rotators (0.04$<$R$_o<$0.4) as fast (R$_o<$0.04) or slow (R$_o>$0.44) rotators; this may be astrophysical or a result of period-detection sensitivity. We discover 30 fast, 59 intermediate, and 33 slow rotators. We measure a median starspot coverage of 13\% of the stellar hemisphere and constrain the minimum magnetic field strength consistent with our flare energies and spot coverage to be 500 G, with later-type stars exhibiting lower values than earlier-types. We observe a possible change in superflare rates at intermediate periods. However, we do not conclusively confirm the increased activity of intermediate rotators seen in previous studies. We split all rotators at R$_o \sim$0.2 into P$_{Rot}<$10 d and P$_{Rot}>$10 d bins to confirm short-period rotators exhibit higher superflare rates, larger flare energies, and higher starspot coverage than do long-period rotators, at p-values of 3.2$\times$10$^{-5}$, 1.0$\times$10$^{-5}$ and 0.01, respectively.
\end{abstract}


\maketitle

\section{Introduction}
Stellar rotation and surface magnetic activity (e.g. surface field topology, starspots, and flares) are intrinsically related phenomena. Quickly-rotating young stars drive increased surface magnetic activity, while surface magnetism controls the spin-down of stellar rotation with age \citep{Weber_1967,Kawaler1988}. Spin-down from angular momentum loss depends on the coupling of the field to the stellar wind, with complex fields resulting in orders-of-magnitude weaker coupling than dipole-dominant fields (e.g. \citealt{Brown2014,Garraffo2015,Garraffo2016_b,Garraffo2018}). Directly measuring whether surface field topology is simple or complex is difficult and expensive, and has only been performed in detail for about 10$^2$ cool stars with well-constrained stellar rotation periods \citep{Shulyak2017}. However, measuring the surface magnetic activity levels of many stars at a range of rotation periods may indirectly probe magnetic topology throughout spin-down.

\subsection{Stellar activity as a probe of spin-down}\label{intro_large_surveys_rotate_active}

Large surveys of stellar rotation periods provide insight into the periods at which the magnetic field may change from simple to complex topologies. Stellar rotation surveys find few cool stars with rotation periods between ten and seventy days, but many faster and slower rotators (e.g \citet{Newton2016,Newton2018,Oelkers2018}). The transition from the quickly-rotating phase to slowly-rotating phase is therefore thought to occur rapidly for cool stars \citep{Newton2016,Newton2018} due to a change in the state of the surface magnetic field and the sudden increased rate of mass and angular momentum loss (AML) that results \citep{Brown2014}. High mass stars spin down earlier than low mass stars; many field-age M-dwarfs are still actively spinning down \citep{Newton2016}.

Stellar activity (e.g. stellar flaring and starspot coverage) is well known to decrease as stars spin down with age \citep{Ambartsumian1975}. It is hypothesized that increased stellar activity may be observed from cool stars with intermediate rotation periods as the surface magnetic field evolves from a simple into a complex topology \citep{Mondrik2019}. Two common photometric measurements that may trace the evolution of the magnetic field are the sinusoidal oscillations in brightness from starspots and the amount of stellar flaring.

Starspots are often used to measure the stellar rotation period (e.g. \citealt{Baliunas1996,Affer2012,Newton2016,Oelkers2018}). Starspots are a form of stellar activity that appear on the photosphere of a star and are effects of the interior stellar magnetic dynamo. Starspots are cooler than the rest of the photosphere, resulting in a flux difference between the spotted and non-spotted surfaces of a star \citep{Berdyugina2005}. As the photosphere rotates, starspots often induce regular brightness variations in stellar photometry. The fraction of the stellar hemisphere covered by starspots, or starspot coverage fraction, decreases at long rotation periods for stars above the fully-convective mass limit, probing the evolution of the star's surface magnetic field throughout spin-down (e.g. \citealt{Hartman2011,McQuillan2014,Newton2016,Notsu2019,Morris2019}).

Stellar flares are another indicator of surface magnetism. Flares occur when the surface magnetic field re-connects, impulsively releasing electromagnetic radiation. Cool stars are often flare stars, even emitting frequent superflares (e.g. \citealt{Paudel2018, Howard2019}): extremely intense flares that release 10-1000X more energy than those seen from the Sun. As M-dwarfs age, both the flare occurrence rate and flare energy decrease \citep{Davenport2019,Ilin2019}. Flaring also depends upon stellar mass due to the depth of the convective layer \citep{davenport2016}. Because flares are intimately connected with the surface field and depend on stellar rotation, it is hypothesized they may be useful in separating M-dwarfs with complex and simple fields. An increased flare rate from late M-dwarfs has been observed at intermediate rotation periods (10$<P_\mathrm{Rot}<$70 d), supporting this hypothesis \citep{Mondrik2019}.

Starspot coverage and flaring are closely linked. The largest flare a star may emit is limited by the stored magnetic energy of the starspot group that produced it. By comparing the largest flare observed from each star and the starspot coverage fraction of that star, the stellar magnetic field strength may be constrained. This is because the surface magnetic field strength adjusts the conversion from starspot size to flare energy; the field must allow the observed flares given the observed spot sizes \citet{Notsu2019}. Similarly, estimates may be made for the surface magnetic field strengths of cool rotators as they spin down. Combining a large sample of stellar flares and rotation periods allows estimates of their minimum surface magnetic fields to be tested against typical magnetic field strengths of cool stars \citep{Shulyak2017}.

\subsection{Photometric surveys of rotating cool stars}\label{intro_survey_comparison}

Large numbers of photometric rotation periods of cool stars have been or are being catalogued by various space-based and ground-based surveys. Examples include 5257 \textit{Kepler} \citep{Borucki2010} K5 and later rotators (at least 80\% of which are M1 or earlier) from \citet{McQuillan2014}, at least 10$^5$-10$^6$ K5 and later rotators from the Transiting Exoplanet Survey Satellite (TESS; \citealt{Ricker_TESS}) estimated from \citet{Stassun2019,Zhan2019}, $\sim$800 K5 or later rotators in the Kilodegree Extremely Little Telescope (KELT; \citealt{Pepper2003,Pepper2007,Pepper2012}) from \citet{Oelkers2018}, and 628 mid-to-late M-dwarf rotators from MEarth \citep{Nutzman2008,Berta2012} \citep{Newton2016,Newton2018}. While about 1-10\% of late K-dwarfs and early M-dwarfs are flare stars, about 30\% of mid-to-late M-dwarfs are flare stars (e.g. \citealt{Yang2017,gunter2019,Howard2019}). Cross-matching stars in each survey with rotation periods against stars with stellar flares therefore significantly reduces the sample size.

\subsection{Stellar rotation with TESS and Evryscope}\label{intro_Evryscope}
Optimized to observe low-amplitude variation from all nearby cool stars, TESS will contribute the majority of fast and intermediate-period cool rotators. However, the TESS primary mission observes most stars for only 28 days, decreasing its ability to measure the periods of slow rotators. Furthermore, the uncertainties to the periods of intermediate and slow rotators obtained by TESS will be large (e.g. errors from approximately $\sim$0.1-to-$\sim$1 days) compared to longer-duration observations.

The Evryscope \citep{Evryscope2015,Ratzloff2019} observes all bright cool stars across the Southern sky. The Evryscope is an array of small telescopes which simultaneously image the entire accessible sky, producing light curves of all ($\sim$0.5$\times$10$^5$) nearby cool stars. Evryscope light curves allow detection of significantly longer rotation periods than from TESS data alone. While TESS observes each star for $\sim$28 days in the red at high photometric precision, Evryscope observes each star at moderate precision for several years in the blue. Combined rotation periods in the blue and in the red allows not only better error analysis of the rotation rate for large numbers of field stars during spin-down, but also an estimate of the color-dependence of starspot modulation during this process. Long-term monitoring by Evryscope also confirms whether periodic brightness modulation seen in TESS is transient or stable over the course of multiple years to better inform RV follow-up efforts of planet candidates.

In this work, we focus on the subset of Evryscope rotation periods of previously-identified flare stars from \citet{Howard2019}. This subset of the Evryscope data was selected from cool stars with 2-minute cadence light curves from both Evryscope and TESS, allowing a comparison of Evryscope and TESS rotation. Future work will further explore the combined flare rate and starspot coverage in both the TESS and Evryscope bands.

In Section \ref{EvryFlare} of this work, we describe the Evryscope, light curve generation, and rotation period, starspot, and stellar flare observations. We also describe the TESS observations. In Section \ref{evr_p_rot_measurements}, we describe rotation period detection in Evryscope and TESS and estimation of period uncertainties. We describe how the sinusoidal amplitude of rotation is greater in the Evryscope $g^{\prime}$ bandpass than in the red TESS bandpass and how this effect is greatest for low-mass stars. In Section \ref{results}, we describe the distributions of rotation periods, Rossby numbers, amplitudes of sinusoidal rotation, starspot coverage fractions, and surface magnetic field constraints. We discuss the decrease in activity with rotation period, and describe a possible increase in superflare rates at intermediate rotation periods. In Section \ref{discuss_conclude}, we summarize our results and conclude.

\section{EvryFlare: all-sky stellar activity search}\label{EvryFlare}
The EvryFlare survey is an ongoing comprehensive survey of stellar activity from all cool stars observed by Evryscope in the accessible Southern sky. Evryscope monitors large flares, stellar rotation periods, and starspot coverage from all nearby cool stars. 

\subsection{Evryscope observations}\label{evryscope_observations}
As part of the Evryscope survey of all bright Southern stars, we discover many variable stars and rotating stars with starspots. The Evryscope is an array of small telescopes that simultaneously images 8150 square degrees and 18,400 square degrees in total each night on the sky. Evryscope observes at two-minute cadence in \textit{g}\textsuperscript{$\prime$}~\citep{Evryscope2015}, and is optimized for bright, nearby stars, with a typical dark-sky limiting magnitude of \textit{g}\textsuperscript{$\prime$}=16. Each night, Evryscope continuously monitors each part of the sky down to an airmass of two and at a resolution of 13\arcsec pixel$^{-1}$ for $\sim$6 hours, The system accomplishes this by employing a ``ratchet" strategy, tracking the sky for 2 hours at a time before ratcheting back into the initial position and continuing observations \citep{Ratzloff2019}.

The Evryscope has already obtained 3.0 million raw images, which we store as $\sim$250~TB of data. Evryscope images are reduced at real-time rates using a custom data reduction pipeline \citep{Law2016,Ratzloff2019}. Each 28.8 MPix Evryscope image is calibrated using a custom wide-field astrometric solution algorithm. Background modeling and subtraction is carefully performed before raw photometry is extracted within forced-apertures at coordinates in an Evryscope catalog of 3M known source positions, including all stars brighter than \textit{g}\textsuperscript{$\prime$}=15, fainter cool stars, white dwarfs, and a variety of other targets. We then generate light curves across the Southern sky by differential photometry in small sky regions using carefully-selected reference stars and across several apertures \citep{Ratzloff2019}. Systematics are partially removed by employing two iterations of the SysRem detrending algorithm \citep{tamuz2005}.

We periodically regenerate the entire database of Evryscope light curves in order to incorporate observations obtained since the last update and to improve the photometric precision. At the time the data was analyzed for the present work, the Evryscope light curve database spanned Jan 2016 through June 2018, averaging 32,000 epochs per star (with factors of several increases to this number closer to the South Celestial Pole). Depending upon the level of stellar crowding, light curves of bright stars (\textit{g}\textsuperscript{$\prime$}=10) reach 6 mmag to 1\% photometric precision. Evryscope light curves of dim stars (\textit{g}\textsuperscript{$\prime$}=15) reach comparable precision to TESS, attaining 10\% photometric precision \citep{Ratzloff2019}.

\subsection{TESS observations}\label{tess_observations}
The TESS mission is searching for transiting exoplanets across the entire sky, split into 26 sectors. TESS observes each sector continuously in the red with four 10.5 cm optical telescopes for 28 days at 21$\arcsec$ pixel$^{-1}$. We chose our original sample to have calibrated, short-cadence TESS light curves during the Primary Mission, which were downloaded from MAST\footnote{https://mast.stsci.edu} for each cool flare star in our sample. We selected Simple Aperture Photometry (SAP) light curves rather than Pre-search Data Conditioning (PDC) ones to avoid removing real astrophysical variability.

\subsection{Evryscope+TESS sample of cool flaring rotators}\label{evryscope_sample}
We search for rotation periods in our sample of flaring cool stars (i.e. K5-M4 dwarfs) from \citet{Howard2019}. Although Evryscope observes $\sim$0.5$\times$10$^5$ cool stars, 2-minute cadence light curves of only 4,068 cool stars were produced by both Evryscope and TESS in the first six TESS sectors. We selected only stars with both a high-cadence light curve in the blue (Evryscope) and in the red (TESS) in order to compare the flare amplitudes, flare energies, flare rates, rotation periods and amplitudes of rotation between these bands. Evryscope observed 575 large flares with a median energy of 10$^{34}$ erg from the 284 flare stars. Of these, rotation was detected for 122 stars. These stars comprise the sample of active cool rotators in this work. Future work will explore a larger sample in both Evryscope and TESS.

The stellar flares were observed in the Evryscope light curves from the subset of rotators within the \citet{Howard2019} sample. These rotators are given here in Table \ref{table:rotation_per_tab}. Flares are discovered and characterized as described in \citet{Howard2019}. Briefly, we searched 2-minute cadence Evryscope light curves for large flares first by eye, and then with the Auto-ELFS automated flare-search algorithm. The algorithm applies a flare matched-filter to the light curve and records brightening events that exceed the local noise by at least 4.5$\sigma$ as flare candidates. Event start and stop times are determined by the first and last epochs to exceed the noise by 1$\sigma$ around the peak epoch. The light curve of each flare candidate is converted to fractional-flux $\Delta$F/F using the out-of-flare flux $F_0$: $\Delta$F/F=$\frac{F-F_0}{F_0}$. The equivalent duration (ED) of each flare candidate is computed from the start to the stop time in seconds by a trapezoidal integration of the fractional-flux. We multiply the ED by the $g^{\prime}$ stellar quiescent flux (L$_0$) computed from the APASS DR9 \citep{Henden2016} $g^{\prime}$ magnitude and Gaia DR2 \citep{Gaia2016,Gaia2018} distance; L$_0$ is given in units of erg second$^{-1}$. Finally, we convert flare energies in $g^{\prime}$ to bolometric energies assuming a 9000 K flare blackbody. These events are inspected by eye for systematics or astrophysics other than flares as described in \citet{Howard2019} and subsequently confirmed or rejected.

We calculated the maximum-energy flare observed from each star during 2+ years of Evryscope observations, as well as the annual superflare rate of each star. We use these two flare star parameters to investigate the dependence of flaring upon stellar rotation and starspot coverage to avoid discovering random correlations between a large number of flaring variables.

\subsection{Characterizing Stellar Properties}\label{stellar_astrophysics}
Obtaining accurate values of stellar effective temperature and stellar radius helps constrain the physical parameters of starspots. All values are given for each star in Table \ref{table:rotation_per_tab}.

\subsubsection{Estimating Photometric Spectral Type}\label{est_spt_type}
We estimate the photometric spectral type of each star. Because \citet{Howard2019} estimated spectral type from one color and a Gaia DR2 distance, we find the \textit{g}$^{\prime}$ blue band may over-predict the stellar effective temperature of cool dwarfs by several hundred K compared to classifiers that use several colors (e.g. \citealt{Ratzloff2019_polar}). To provide increased accuracy in our sub-type classification, we use the photometric spectral type classifier described in \citet{Ratzloff2019_polar}. 
  
Briefly, \citet{Ratzloff2019_polar} classifies main sequence dwarfs by their reduced proper motion (RPM) and multiple stellar colors using a Gaussian Mixture Model (GMM; \citealt{Pedregosa2012arXiv}). The GMM calculates the negative-log-likelihood and confidence level each star has been correctly classified. The GMM classifies M-dwarfs to within at least 3 spectral sub-types 95\% of the time. While it is possible for an RPM classifier to fail to separate dwarf and giant stars at low RPM, we do not consider this to be a concern because the entire sample of stars was separately classified on the basis of Gaia DR2 parallax and APASS DR9 g-magnitude; we desire to increase the precision of sub-type measurements made from one color toward several colors. Out of 122 stars, the GMM classified 80\% of our sample. For the other 20\%, no GMM classification was given, likely a result of having too few cross-matched colors. For stars without a classification, we assign the spectral type via the absolute $g^{\prime}$ magnitude as described in \citet{Howard2019}.
 
\subsubsection{Estimating Stellar Effective Temperature, Mass, and Radius}\label{est_t_eff}
We compute stellar \textbf{effective temperature} from the estimated spectral type using the relations given in Table 5 in the Appendix of \citet{Kraus2007}. We also compute stellar \textbf{mass} from the estimated spectral type using the relations given in Table 5 in the Appendix of \citet{Kraus2007}. We compute stellar \textbf{radius} using the mass-radius relationship for cool stars given in \citet{Mann2015}.
  
\subsubsection{Characterizing Starspots and Flares}\label{starspot_parameters}
Assuming that the observed sinusoidal stellar brightness variations are caused by star spots rotating into and out of view, we may investigate the nature of the star spots in our sample. For each rotating star in our sample, we may estimate the following starspot parameters:

\begin{itemize}
  \item We estimate starspot temperature using the relationship between stellar effective temperature T$_\mathrm{Eff}$ and starspot temperature T$_\mathrm{Spot}$ from \citet{Notsu2019}: 
  \begin{equation}
    \begin{aligned}
      \Delta T(T_\mathrm{Star})=T_\mathrm{Star} - T_\mathrm{Spot}=\\
      3.58 \times 10^{-5} T_\mathrm{Star}^2 + 0.249 T_\mathrm{Star}-808
    \end{aligned}
    \label{eq:delta_spot_temp}
  \end{equation}
  We note this fit was derived for solar type stars observed by \textit{Kepler} and is extrapolated into the cool star regime. We therefore urge caution in the application of these values.
  \item We measure \textbf{spot coverage} as the starspot area A$_\mathrm{Spot}$ divided by the projected hemispherical area of the star A$_\mathrm{Star}$. We use the relation described in \citet{Maehara2012,Shibata2013,Notsu2013, Notsu2019}: 
  \begin{equation}
    \begin{aligned}
      \frac{A_\mathrm{Spot}}{A_\mathrm{Star}} = \frac{\Delta F}{F} \left[1-\left(\frac{T_\mathrm{Spot}}{T_\mathrm{Star}}\right)^{4}\right]^{-1}
    \end{aligned}
    \label{eq:spot_cover_fraction}
  \end{equation}
  $\Delta$F/F is the normalized flux difference in brightness between the brightest part of the star and the dimmest side and is in units of fractional-flux. A$_\mathrm{star}$ is given as A$_\mathrm{Star}=\pi R_\mathrm{Star}^{2}$. The bolometric spot area will differ from the spot area measured in a given bandpass.
\end{itemize}

\begin{figure*}
	\centering
	{
		\includegraphics[trim= 1 1 1 1,clip, width=6.9in]{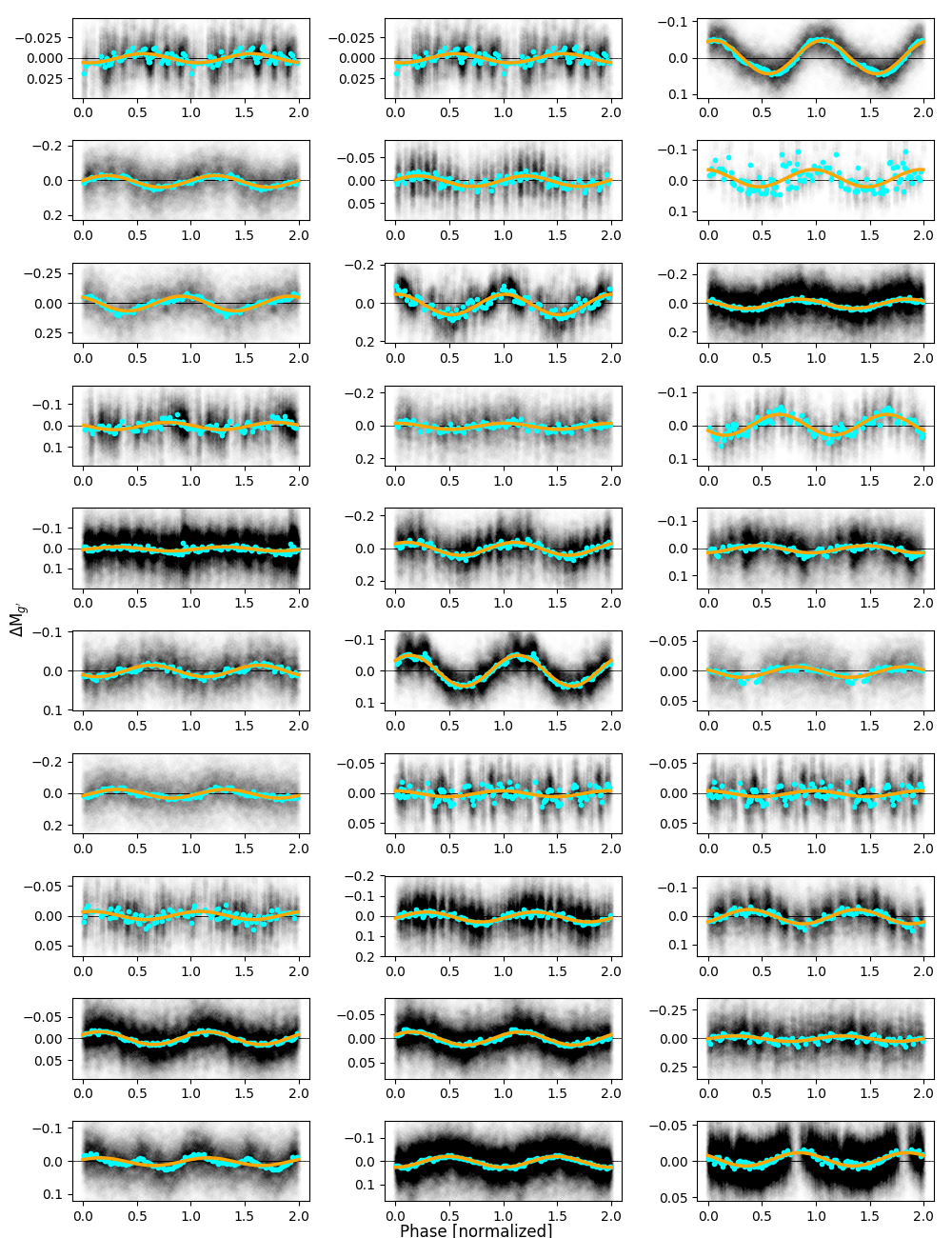}
	}
	\caption{A random subset of all photometric rotation periods found in Evryscope light curves for 122 cool flare stars. In each panel, we plot $\Delta$M$_{g'}$ magnitudes versus phase. We repeat the phased epochs twice to better display the periodicity. A phased and binned Evryscope light curve is overlaid (in blue), along with a best-sinusoid fit to the unbinned data (in orange). We sometimes detect periods with additional periodicity at harmonics of the strongest peak, such as in the bottom-left panel.}
	\label{fig:grid_evr_prots}
\end{figure*}

\begin{figure*}
	\centering
	{
		\includegraphics[trim= 1 1 1 1,clip, width=6.9in]{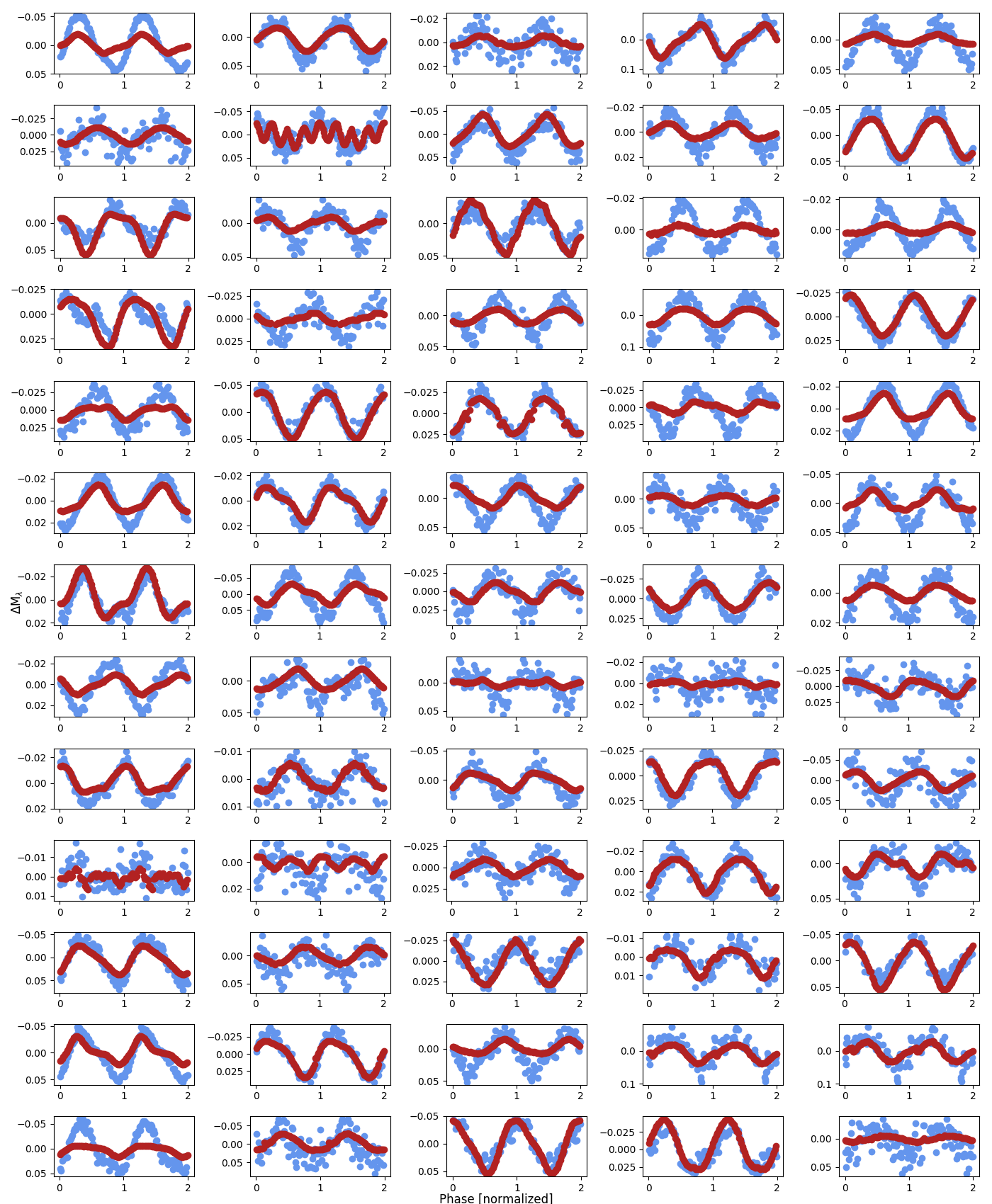}
	}
	\caption{Phased and binned light curves of a subset of cool rotators for which the TESS light curve folds up exactly to the Evryscope-detected period. The phased and binned Evryscope (blue) and TESS (red) light curves are overlaid. In each panel, we plot the normalized flux $\Delta$F/F versus phase. We repeat the phased epochs twice to better display the periodicity. We find the amplitudes of the TESS light curves are almost universally less than or equal to the Evryscope amplitudes. We note the increase in spot contrast in the blue $g^{\prime}$ bandpass versus the red $T$ bandpass. TESS amplitudes are further decreased beyond the initial amplitude difference by systematics-removal. In visual inspection and A-D tests, this color difference correlates with the stellar effective temperature of our K5-M4 stars but not with the presence of companion stars in the TESS pixel, which is 4$\times$ larger.}
	\label{fig:grid_evr_vs_TESS_phased}
\end{figure*}

These results and relevant uncertainties are displayed in Table \ref{table:rotation_per_tab} for each flare star. The measured rotation period and period error calculated as described in Section \ref{evr_p_rot_measurements} is also included for each rotating flare star. We plot a grid of sample Evryscope period detections in Figure \ref{fig:grid_evr_prots}. We also plot a grid of sample Evryscope and TESS period detections overlaid on each other in Figure \ref{fig:grid_evr_vs_TESS_phased}.

\begin{turnpage}

\begin{table*}

\caption{Rotation Periods and Starspots on Cool Flare Stars Observed by Evryscope in TESS Sectors 1-6}
\begin{tabular}{p{1.4cm} p{0.8cm} p{1.3cm} p{1.2cm} p{0.9cm} p{0.8cm} p{0.5cm} p{0.5cm} p{1.3cm} p{0.7cm} p{0.7cm} p{0.8cm} p{0.8cm} p{0.8cm} p{0.9cm} p{0.7cm} p{0.7cm} p{0.6cm} p{0.6cm} p{0.6cm}}
\hline
 &  &  &  &  &  &  &  &  &  &  &  &  &  &  &  &  &  &  \\
TIC ID & Sector & RA & Dec & P$_\mathrm{Rot}$ & P$_\mathrm{Rot}$ Err. & R$_o$ & LS \newline S/N & Time of \newline  peak phase & EVR \newline ampl. rot. & EVR frac. light blocked & Rot. in \newline TESS? & TESS \newline ampl. rot. & TESS frac. light blocked & Spot temp. & EVR spot \newline cover & Mass & Rad. & T$_\mathrm{Eff}$ & SpT \\
 & & [J2018] & [J2018] & [d] & [d] & & [$\sigma$] & [MJD] &  [$\Delta$M$_{\textit{g}\textsuperscript{$\prime$}}$] & [$\Delta$F/F] &  & [$\Delta$M$_{T}$] & [$\Delta$F/F] & [K] &  & [M$_{\odot}$] & [R$_{\odot}$] & [K] & \\
\hline
 &  &  &  &  &  &  &  &  &  &  &  &  &  &  &  &  &  &  & \\
 &  &  &  &  &  &  &  & ... &  &  &  &  &  &  &  &  &  &  & \\
 &  &  &  &  &  &  &  &  &  &  &  &  &  &  &  &  &  &  & \\
44796808 & 4+5 & 62.41391 & -26.8143 & 3.74 & 0.03 & 0.06 & 11 & 57401.102 & 0.04 & 0.075 & yes & 0.015 & 0.028 & 2900 & 0.18 & 0.29 & 0.31 & 3400 & M3 \\ 
50745582 & 6 & 83.01872 & -3.0913 & 4.3742 & 0.0006 & 0.1 & 27 & 58159.047 & 0.045 & 0.084 & yes & 0.036 & 0.067 & 3000 & 0.18 & 0.42 & 0.42 & 3500 & M2 \\ 
55368621 & 1-6 & 77.4463 & -60.0016 & 45.1 & 0.1 & 1.65 & 16 & 57699.241 & 0.026 & 0.048 & no &   &   & 3300 & 0.08 & 0.63 & 0.59 & 4100 & K7 \\ 
71335815 & 1 & 331.21484 & -47.5337 & 9.534 & 0.002 & 0.22 & 28 & 57610.053 & 0.054 & 0.101 & yes & 0.036 & 0.066 & 3000 & 0.22 & 0.42 & 0.42 & 3500 & M2 \\ 
77957301 & 5+6 & 79.34437 & -35.3657 & 1.925 & 0.009 & 0.04 & 11 & 57746.101 & 0.017 & 0.031 & prob &   &   & 3000 & 0.07 & 0.36 & 0.37 & 3400 & M2.5 \\ 
77959676 & 5+6 & 79.62111 & -30.0256 & 1.6961 & 0.0002 & 0.06 & 25 & 57829.069 & 0.021 & 0.039 & yes & 0.006 & 0.011 & 3200 & 0.07 & 0.59 & 0.56 & 3800 & M0 \\ 
79566479 & 3+4 & 52.98208 & -43.9871 & 2.9207 & 0.0003 & 0.1 & 27 & 57639.331 & 0.054 & 0.101 & yes & 0.038 & 0.071 & 3200 & 0.18 & 0.62 & 0.58 & 4000 & K8 \\ 
88479623 & 4 & 47.02903 & -24.7596 & 0.918 & 0.001 & 0.02 & 23 & 57698.159 & 0.014 & 0.026 & harm & 0.019 & 0.035 & 3000 & 0.05 & 0.48 & 0.47 & 3600 & M1.5 \\ 
89205615 & 4 & 58.09812 & -28.4386 & 0.3487 & 0.0001 & 0.01 & 13 & 57700.241 & 0.044 & 0.082 & yes & 0.038 & 0.072 & 3100 & 0.16 & 0.54 & 0.52 & 3700 & M1 \\ 
115242300 & 2 & 6.26900 & -36.7714 & 12.7 & 0.007 & 0.2 & 19 & 57662.214 & 0.032 & 0.06 & yes & 0.011 & 0.020 & 2900 & 0.14 & 0.29 & 0.31 & 3400 & M3 \\ 
140045537 & 1 & 329.41269 & -51.0094 & 1.1 & 0.1 & 0.01 & 34 & 57613.110 & 0.019 & 0.035 & yes & 0.006 & 0.011 & 2800 & 0.10 & 0.20 & 0.23 & 3200 & M4 \\ 
140460192 & 3 & 20.51881 & -33.6176 & 10.0 & 0.2 & 0.23 & 31 & 57953.418 & 0.031 & 0.058 & beat & 0.020 & 0.036 & 3000 & 0.12 & 0.42 & 0.42 & 3500 & M2 \\ 
200364466 & 4+5+6 & 77.26495 & -42.1553 & 1.605 & 0.007 & 0.03 & 8 & 57418.083 & 0.067 & 0.127 & no &   &   & 3000 & 0.29 & 0.36 & 0.37 & 3400 & M2.5 \\ 
200368439 & 5+6 & 77.46538 & -42.1288 & 3.6281 & 0.0005 & 0.1 & 31 & 57398.144 & 0.072 & 0.136 & yes & 0.026 & 0.048 & 3000 & 0.28 & 0.48 & 0.47 & 3600 & M1.5 \\ 
201426753 & 6 & 92.3298 & -35.8254 & 1.7177 & 6e-05 & 0.04 & 28 & 57682.321 & 0.025 & 0.046 & yes & 0.022 & 0.041 & 3000 & 0.10 & 0.42 & 0.42 & 3500 & M2 \\ 
201919099 & 2+3 & 40.63832 & -57.6606 & 7.373 & 0.002 & 0.26 & 28 & 57752.082 & 0.053 & 0.1 & yes & 0.044 & 0.083 & 3200 & 0.18 & 0.62 & 0.58 & 4000 & K8 \\ 
206327795 & 1 & 354.07498 & -48.5878 & 7.9 & 0.2 & 0.25 & 11 & 58012.294 & 0.008 & 0.015 & no &   &   & 3100 & 0.03 & 0.56 & 0.54 & 3800 & M0.5 \\ 
206556127 & 1 & 333.66052 & -21.6977 & 2.2044 & 7e-05 & 0.07 & 22 & 57936.264 & 0.034 & 0.063 & yes & 0.022 & 0.040 & 3100 & 0.12 & 0.56 & 0.54 & 3800 & M0.5 \\ 
207199350 & 2+3+4 & 47.43713 & -57.5494 & 5.95 & 0.1 & 0.16 & 25 & 58077.234 & 0.035 & 0.065 & yes & 0.014 & 0.027 & 3000 & 0.13 & 0.48 & 0.47 & 3600 & M1.5 \\ 
219315573 & 1 & 320.30832 & -59.4634 & 9.785 & 0.01 & 0.13 & 12 & 57934.349 & 0.088 & 0.168 & no &   &   & 2900 & 0.42 & 0.24 & 0.27 & 3300 & M3.5 \\ 
229807000 & 1 & 352.24075 & -68.043 & 0.3746 & 0.0002 & 0.01 & 29 & 57905.243 & 0.022 & 0.041 & yes & 0.020 & 0.037 & 3000 & 0.09 & 0.36 & 0.37 & 3400 & M2.5 \\ 
231020638 & 2+3 & 27.73764 & -58.7343 & 1.6398 & 0.0002 & 0.02 & 18 & 57635.093 & 0.048 & 0.09 & yes & 0.018 & 0.032 & 2800 & 0.24 & 0.20 & 0.23 & 3200 & M4 \\ 
231799463 & 4+5+6 & 78.25721 & -70.4581 & 2.1392 & 0.0002 & 0.05 & 10 & 57834.018 & 0.019 & 0.035 & prob &   &   & 3000 & 0.08 & 0.42 & 0.42 & 3500 & M2 \\ 
231835378 & 1+2 & 24.42091 & -64.4495 & 9.584 & 0.002 & 0.15 & 14 & 57671.204 & 0.07 & 0.133 & no &   &   & 2900 & 0.31 & 0.29 & 0.31 & 3400 & M3 \\ 
231867117 & 1+2 & 6.03841 & -62.1845 & 1.7498 & 0.0001 & 0.03 & 23 & 58022.233 & 0.026 & 0.048 & yes & 0.027 & 0.049 & 3000 & 0.11 & 0.36 & 0.37 & 3400 & M2.5 \\ 
232077453 & 1+2 & 29.44500 & -67.6336 & 29.4 & 1.5 & 0.46 & 21 & 58139.077 & 0.057 & 0.107 & harm & 0.004 & 0.008 & 2900 & 0.25 & 0.29 & 0.31 & 3400 & M3 \\ 
471016669 & 2 & 359.33569 & -12.9800 & 7.627 & 0.002 & 0.09 & 12 & 57633.150 & 0.03 & 0.056 & yes & 0.008 & 0.015 & 2800 & 0.15 & 0.20 & 0.23 & 3200 & M4 \\ 
 &  &  &  &  &  &  &  & ... &  &  &  &  &  &  &  &  &  &  & \\
 &  &  &  &  &  &  &  &  &  &  &  &  &  &  &  &  &  &  & \\
\hline
\end{tabular}
\label{table:rotation_per_tab}
{\newline\newline \textbf{Notes.} Parameters of 122 rotating flare stars monitored by Evryscope (1 star per row). This is a subset of the full table. The full table is available in machine-readable form. The columns are: TIC ID, the TESS sector(s) the star was observed (if observed for more than 3 but less than six of the sectors, ``most" is recorded), RA and Dec (the current Evryscope-measured positions of the star), the stellar rotation period in days, the uncertainty on the period in days, the Rossby number, the signal over noise of the Lomb-Scargle peak, a time of peak rotational brightness in the phase-folded Evryscope light curve, the Evryscope-measured sinusoidal amplitude of rotation in $\Delta$M$_{\textit{g}\textsuperscript{$\prime$}}$, the Evryscope-measured double amplitude of rotation in fractional-flux (i.e. the fraction of starlight blocked by spots), a note whether the rotation period observed by Evryscope is also visible in the TESS light curve (choices are yes for yes, no for no, harm. for a harmonic period, beat for a beat frequency of 1 d and the TESS period, prob. for a likely but noisy match, long for too long to observe in TESS), the TESS-measured amplitude of rotation in $\Delta$M$_{TESS}$, the TESS-measured double-amplitude of rotation in fractional-flux (i.e. the fraction of starlight blocked by spots), the estimated starspot temperature in K, the Evryscope-measured starspot coverage fraction, the estimated stellar mass in M$_{sol}$, the estimated stellar radius in R$_{sol}$, the estimated stellar effective temperature in K, and the estimated spectral type.}
\end{table*}

\end{turnpage}

\section{Evryscope rotation periods}\label{evr_p_rot_measurements}
We search for photometric rotation periods by computing the Lomb-Scargle (LS) periodogram \citep{Lomb1976,Scargle1982,VanderPlas2018} of each Evryscope light curve.

\subsection{Initial detection of periods in Evryscope}\label{initial_period_search}

\begin{figure}
	\centering
	{
		\includegraphics[trim= 1 1 1 1,clip, width=3.4in]{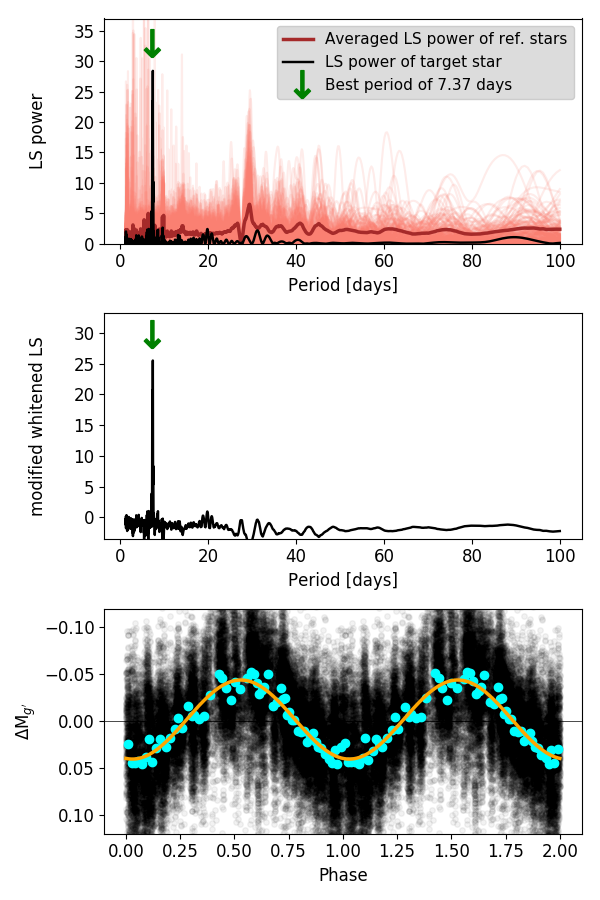}
	}
	\caption{An example photometric rotation period found in an Evryscope light curve. The LS periodograms of all stars are plotted on top of each other in a transparent red color, while the ``averaged" periodogram is plotted as a solid dark red line. The LS periodogram of the target star is plotted as a solid black line. The averaged LS periodogram is then subtracted from the LS periodogram of the target star and searched for the highest peak above the noise, as displayed in the middle panel of Figure \ref{fig:example_LS_detection}. The best period is denoted by a green arrow. In the bottom panel, we plot $\Delta$M$_{g'}$ magnitudes versus phase. A folded and binned Evryscope light curve is plotted in blue points and compared to the best-fit sine in orange.}
	\label{fig:example_LS_detection}
\end{figure}

We separately compute the LS periodogram of each light curve for 10,000 frequency steps over a test period range of 0.1 to 3 days, and for 10,000 frequency steps over a test period range of 1.25 to 100 d. Periodograms were computed separately in these period ranges as part of a modular data analysis and then the clearest candidate signals across both periodograms were selected. This was a result of realizing the initial lower period limit of 1.25 d had removed fast rotators from the sample. The upper limit of 10$^2$ is arbitrarily selected; we note most active stars will rotate much faster. We also note that distinguishing signal from our systematics and noise becomes increasingly difficult at very long periods, placing us in a different regime for rotation than MEarth \citep{Newton2018}, etc. We also note that if the highest LS peak for a star is at 100 d, we manually increase the period in steps of 0.1 to 0.5 d and examine the phase-folded light curve to see if the true period is slightly larger than 100 d. We subtract 27.5 day and 1 day best-fit sines from all light curves before computing the periodograms. LS power is computed as the median-subtracted LS periodogram peak of the target star over the ``noise" of the periodogram. We exclude a period region within 0.05 days of the detected peak from the noise computation.

In order to constrain systematics during the period analysis, we compare the LS periodogram of each target star with the combined LS periodograms of the other 283 flare stars in \citet{Howard2019}, stepping through the entire sample star by star. Systematic behavior common to all light curves will increase the LS power of each star at systematics-affected periods. We therefore combine together the LS periodograms of all rotating and non-rotating stars, computing the median and standard deviation of the detected LS powers of all stars at each test period from 0.1 to 100 d. We define the averaged LS periodogram as the 1$\sigma$ upper limit of the distribution of LS powers at each tested period. This process is illustrated in the top panel of Figure \ref{fig:example_LS_detection}. We subtract the averaged LS periodogram from the target star periodogram to produce a ``modified pre-whitened (MP) periodogram." The MP periodogram allows the detection of high-amplitude astrophysical oscillations at periods that may also display low amplitude systematic periods while removing the low-amplitude events. For such high-amplitude signals, the height of the peak is reduced in the MP periodogram. Evryscope-detected periods within 5\% of 1 d (or 1/2 d, 1/3 d, 1/4 d, 1/5 d etc) are not considered at all due to the prevalence of the day-night cycle systematic.

The highest peak above the noise in the MP periodogram is selected as the best candidate period as shown in the middle panel of Figure \ref{fig:example_LS_detection}. Candidate periods are investigated in a custom graphical user interface (GUI) by eye; the GUI is an interactive version of Figure \ref{fig:example_LS_detection}. The light curve is folded to the period with the highest peak and visually confirmed as a sinusoid. If the highest peak is not a clear sinusoid, other large peaks above the noise are inspected in the same way. The highest peak is sometimes a harmonic of the true rotation period or even a systematic in the light curve. If a clear sinusoidal signal can be detected, that period is recorded. The light curve of the target star is folded to the best-detected period in the bottom panel of Figure \ref{fig:example_LS_detection}. If the LS power and oscillation amplitude are small and the power spectrum is noisy or dominated by systematic periods, we record no period for that target star. The best estimate for the period of each flare star is recorded in Table I of this work.

\subsection{Bootstrap Measurement of period uncertainty }\label{period_error_boostrap}

A periodogram bootstrap may serve two closely-related purposes: (1) to measure the false-alarm probability of a signal, and (2) to measure the uncertainty of a given period on the data \citep{VanderPlas2018}. We use TESS light curves to assess (1), and use Evryscope light curves and our bootstrap routine to assess (2). While one might initially assume a full test period range of 0.1 to 100 d would best sample the bootstrap uncertainties, a narrower window centered on the detected period provides more meaningful information. A narrow window reduces the effects of aliasing. A 0.1 to 100 d window would result in unphysically-large deviations in the average maximum-power position in the bootstrapped periodograms. The day-night cycle, the lunar cycle, seasonal changes, and instrumental effects will also each imprint on the full periodogram as a convolution of periods \citep{VanderPlas2018}. Therefore, we choose a window size of 20\%, exceeding the FWHM of the detected LS peak. Phase-folding the Evryscope light curve at periods outside the FWHM demonstrates a highly-degraded signal compared to phase-folding at periods within the LS peak.

Uncertainty to each Evryscope-detected period is computed with 1000 trials of a custom bootstrap algorithm, which randomly drops 10\% of the light curve before re-computing the LS periodogram. This method assumes a light curve that is much longer than the oscillation period, and tests if some small section of that light curve may unduly bias the recovered period. In each trial, periodograms are computed with 10,000 steps in frequency within 25\% of the period previously confirmed by eye (chosen to allow up to 20\% error as described below). Periods are tested as follows:
\begin{enumerate}
	\item The bootstrap begins by searching in the periodogram for candidate peaks within 10\% of the period previously confirmed by eye. We start with 10\% of the period to avoid other large peaks in the periodogram that survived the 25\% cut.
	\item If the resulting periods do not converge to better than 10\% (e.g. there are other large peaks in this period range causing the histogram of bootstrapped periods to not be pseudo-normally distributed), the period range of candidate peaks is then extended and the bootstrap is re-run. This time, candidate peaks within 20\% of the period previously confirmed by eye are allowed.
	\item If the resulting periods do not converge to better than 20\%, the bootstrap fails. In this case, the uncertainty to the period is reported to be the FWHM of the LS peak and no further iterations are attempted for that target. Uncertainties larger than 20\% are rare (2\% of the sample) and generally occur only if the period selected by eye that best phases up the light curve is not the highest peak in the periodogram test window.
	\item The final bootstrapped period error is chosen to be the standard deviation of the histogram of bootstrapped values. We ensure at least 250 of the 1000 MC trial values are used in the standard deviation calculation and did not center on another large peak within 25\% of the input period. We also allow for small systematic offsets between the input period (measured by the MP-LS process and not the bootstrap LS) and the distribution of bootstrapped values. When the offset between the input MP-LS period and the median of the bootstrapped period histogram is larger than the standard deviation of the histogram, we increase the error to the larger of the two values. Such offsets are small (3$\sigma$-clipped median of 0.0002 d).
\end{enumerate}
We inspected the bootstrap errors versus the amplitude of rotation to ensure that as amplitudes increase above the photometric noise, the bootstrap errors decrease. This trend loosely holds from amplitudes of 0.008 up to 0.05 mag in $g^{\prime}$. Above 0.05 mag in $g^{\prime}$ the trend of bootstrap error versus amplitude of rotation becomes less clear. Visual confirmation of period errors indicates the smallest errors ($<$10$^{-4}$ d) may be under-estimated.

\subsection{Period validation using TESS light curves}\label{TESS_validation_method}

As a further validation step, we fold the corresponding 2-minute cadence TESS light curve to our detected period. If we observe no coherent behavior at that period in TESS data, we record that information in Table \ref{table:rotation_per_tab}. We note that a lack of TESS periodicity at our detected period does not mean our period is not astrophysical. Starspots evolve over time \citep{Giles2017}, may display a change in contrast against the star at different wavelengths \citep{Notsu2019}, and may even be altered by large flares \citep{Zhan2019}. 

Many TESS SAP flux light curves demonstrate long term trends; to prevent these trends from altering TESS amplitudes of rotation, we pre-whiten the light curves at timescales longer than the Evryscope-detected periodicity. This is done by subtracting a 1D Gaussian-blurred light curve with a blurring kernel equal to the rotation period. We record whether the TESS light curve folds exactly to the Evryscope period in Table \ref{table:rotation_per_tab}. If so, we also record the amplitude of the oscillation in TESS-magnitude and normalized flux in Table \ref{table:rotation_per_tab} for comparison to the Evryscope values. The range of folded TESS light curves that phase to Evryscope periods is visible in Figure \ref{fig:grid_evr_vs_TESS_phased}. 

While folding TESS light curves to the Evryscope period of each rotator identified by eye above, we discovered 27 of our rotation periods in the 1.25+ d range were aliases of an obvious rapid-rotator in TESS. As a result, we re-computed the LS periodogram of all Evryscope light curves down to 0.1 d. For stars with periods already detected in the original 1.25-100 d period search range, we exclude shorter periods at exact beat frequencies of the previously-detected period and 1 d.

We may sometimes detect a period not evidenced in the TESS light curve or vice versa. Systematics in the TESS light curve, in the Evryscope light curve, or in both may cause difficulty in comparing the two periods. In particular, uncorrected TESS systematics in multi-sector light curves may obscure periods of slow rotators.

\newpage
\subsection{Detection of Evryscope periods in TESS}\label{results_evr_periods_in_tess}
During inspection of the TESS light curves of Section \ref{TESS_validation_method}, we observe 75 periods that exactly match in both surveys (shown in Figure \ref{fig:grid_evr_vs_TESS_phased}), and 7 periods that are probably confirmed but do not fold to the exact period detected by Evryscope, possibly due to spot evolution and differential rotation. 4 of our periods appear to be simple harmonics of the fundamental TESS period, and 4 of our periods correspond exactly to the beat frequency of 1 d and the period observed in TESS (all are from the 0.1d to 3 d periodogram). Because astrophysical signals in LS periodograms are well-known to produce power at harmonic frequencies close to the true frequency (i.e. 1/2$\times$, 2$\times$, 1/3$\times$, 3$\times$) \citep{VanderPlas2018}, we include our ``harmonic" and ``beat" detections as genuine detections of the stellar rotation in both surveys. For stars labeled ``harmonic" or ``beat" in Table \ref{table:rotation_per_tab}, we record the unambiguous TESS period. Finally, 3 of our periods are too long to confirm in the TESS light curve. 29 of our periods do not correspond to any period in the TESS light curve.

\subsection{TESS vs. Evryscope sinusoidal amplitudes}\label{tess_ampl}
While folding the high-cadence TESS light curves of each rotator to the Evryscope period as described in Section \ref{evr_p_rot_measurements}, we noticed the Evryscope sinusoidal amplitudes are consistently greater than or equal to those in TESS. We compute the normalized fractional flux difference between Evryscope and TESS amplitudes for the TESS periods of our 75 exact period-matches, 4 harmonic periods, and 4 harmonic-beat periods from Section \ref{high_rates_MC}, for a median flux difference and 1$\sigma$ spread in the distribution of flux difference of 0.04$\substack{+0.03 \\ -0.02}$. It is likely this is an effect of the differing blackbody temperatures of the spot and star. We hypothesized the rotators with the greatest amplitude differences should correlate with stellar effective temperature and therefore color. We checked the correlation visually and observed a weak trend toward larger differences in amplitude at lower masses; we also performed a two-sample Anderson-Darling test on the flux amplitude differences of early and late rotators, and found some correlation (p=0.01, see Section \ref{high_rates_MC} for more information on the test statistic).

To be thorough, we also hypothesized the 4X larger TESS pixels capture more flux from companion stars, diluting the amplitude. We checked the number of Gaia DR2 sources for each star, and found the larger flux differences in amplitude do not correlate with more companion stars (p$\approx$1). We find between 1 and 17 Gaia DR2 sources per 21" aperture; 94\% of our 83 targets have fewer than four nearby sources and display no trend with a difference in flux amplitude. Although not statistically significant, the remaining 6\% of the targets with four or more sources do display flux amplitude differences. We do not see similar amplitude offsets between Evryscope and TESS for other targets (e.g. \citep{Ratzloff2019_EVRCB1, Ratzloff2019_HSDs}) as might be expected if our detection were due to systematics, further supporting our detection of increased contrast with spots at later types.

\section{Discussion: Stellar Activity and Rotation Relations}\label{results}
In this section, we characterize stellar rotation, starspot coverage, and flare energy in the \citet{Howard2019} flare star sample.

\subsection{Stellar rotation periods}\label{results_rot_dist}
\begin{figure}
	\centering
	{
		\includegraphics[trim= 0 1 0 1,clip, width=3.4in]{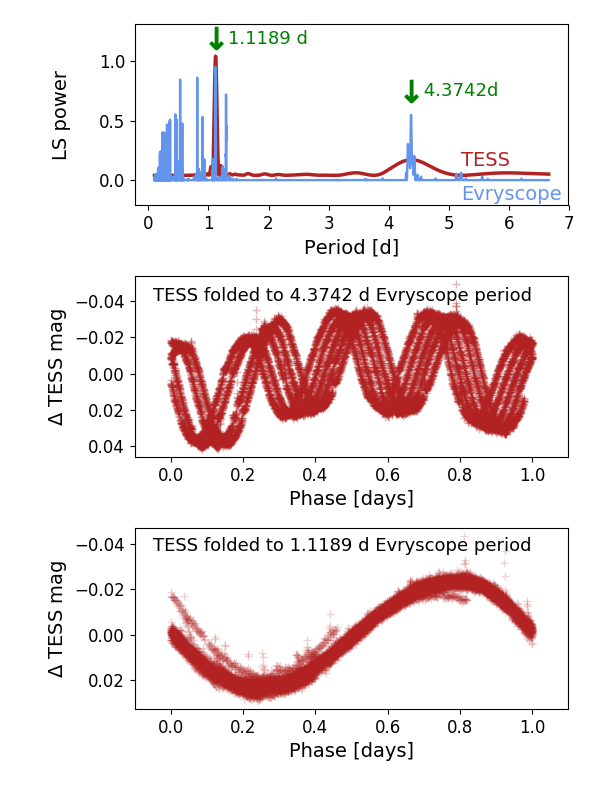}
	}
	\caption{Binarity is observable via multiple rotation periods for the flaring BY Dra variable TIC-50745582 (V* V1311 Ori). Two Evryscope periods were detected and then validated in the TESS light curves. Top panel: The LS periodogram of the TESS light curve and modified pre-whitened LS periodogram of the Evryscope light curve are compared, and the best peaks with $P_\mathrm{rot}<$1.25 d and $P_\mathrm{rot}>$1.25 d are selected, respectively. Bottom panels: The TESS light curve is folded to each period, demonstrating clear rotational modulation.}
	\label{fig:example_binary_periods}
\end{figure}

We discover 122 stellar rotation periods out of 284 flare stars. We detect rotation periods ranging from 0.3487 to 104d, a median and 1$\sigma$ spread of $6.3\substack{+31 \\ -5}$ d. Phase-folded light curves of a random subset of our detected rotation periods are displayed in Figure \ref{fig:grid_evr_prots}. M-dwarf periods of $\sim$7 d are relatively rare in MEarth data, suggesting our sample contains many young stars and stellar binaries. Periods of $\sim$7 d occur when stars are either young and still activity spinning down, or else when they are members of a multiple system that has slowed spin-down \citet{Fleming2019}. Indeed, several stars in the sample are well-known flaring binaries (e.g. GJ 841 A, CC Eridani and V* V1311 Ori, all BY Dra systems \citep{Eker2008,Samus2017}). One way to determine if rotators like these BY Dra are in multiple systems is by multiple periods imprinted on the light curve. Of all our Evryscope rotators, only the BY Dra system V* V1311 Ori clearly showed rotation in both components, as displayed in Figure \ref{fig:example_binary_periods}.

Because the Evryscope light curves are high-cadence and multi-year, many of our detected periods are good to 2-5 significant figures, with better uncertainties for short periods than long periods. The period uncertainties have a median and 1$\sigma$ range of $0.0061\protect\substack{+0.57 \\ -0.0058}$ days. We detect all periods at significance levels greater than 5$\sigma$, with greater significance for shorter periods. The median significance of detection and its 1$\sigma$ range is $18.5\protect\substack{+13 \\ -9}$.

\subsection{Spot Coverage and Maximum Flare Energies}\label{spots_data}
Starspots are easily observed on low-mass stars because the amount of light blocked by spots creates a high-amplitude signal \citep{McQuillan2014}. Starspot coverage fractions are inferred from either the amplitude of rotational modulation in the light curve \citep{Maehara2012,Shibata2013,Notsu2013,Notsu2019}, or comparing TiO bands in stellar spectra with simulated template spectra of the spot and star \citep{Neff1995,ONeal2004,Morris2019}. We search for spots using rotational modulation.  Not all spotted stars will produce photometric rotation periods; rotational variation from spots is suppressed for spots at the poles and stars with spots evenly distributed across the stellar surface \citep{Morris2019}.

\begin{figure*}
	\centering
    \subfigure
	{
		\includegraphics[trim= 5 10 10 0, clip, width=2.25in]{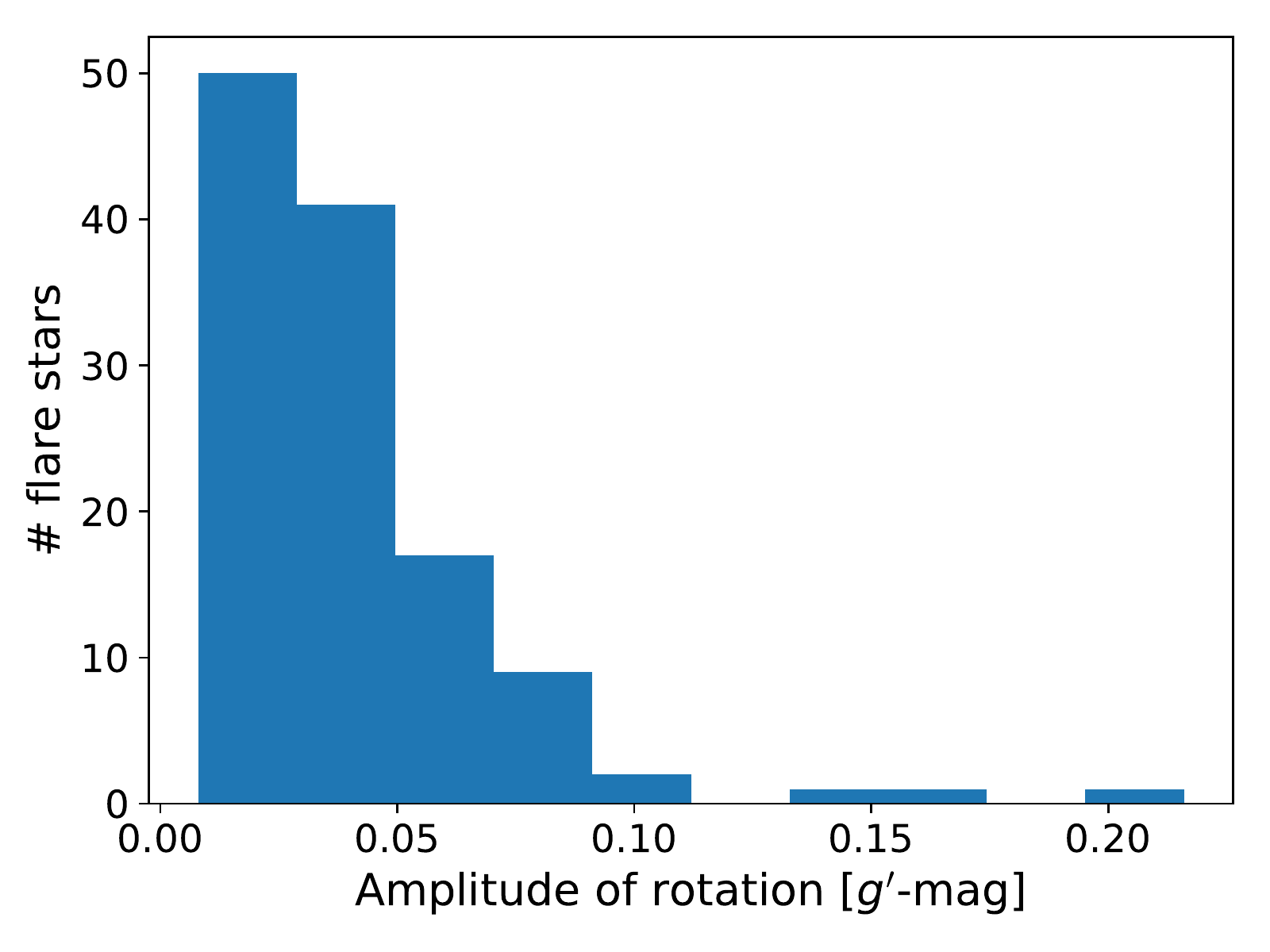}
		\label{fig:spots_overview_a}
	}
	\subfigure
	{
		\includegraphics[trim= 5 10 10 0, clip, width=2.25in]{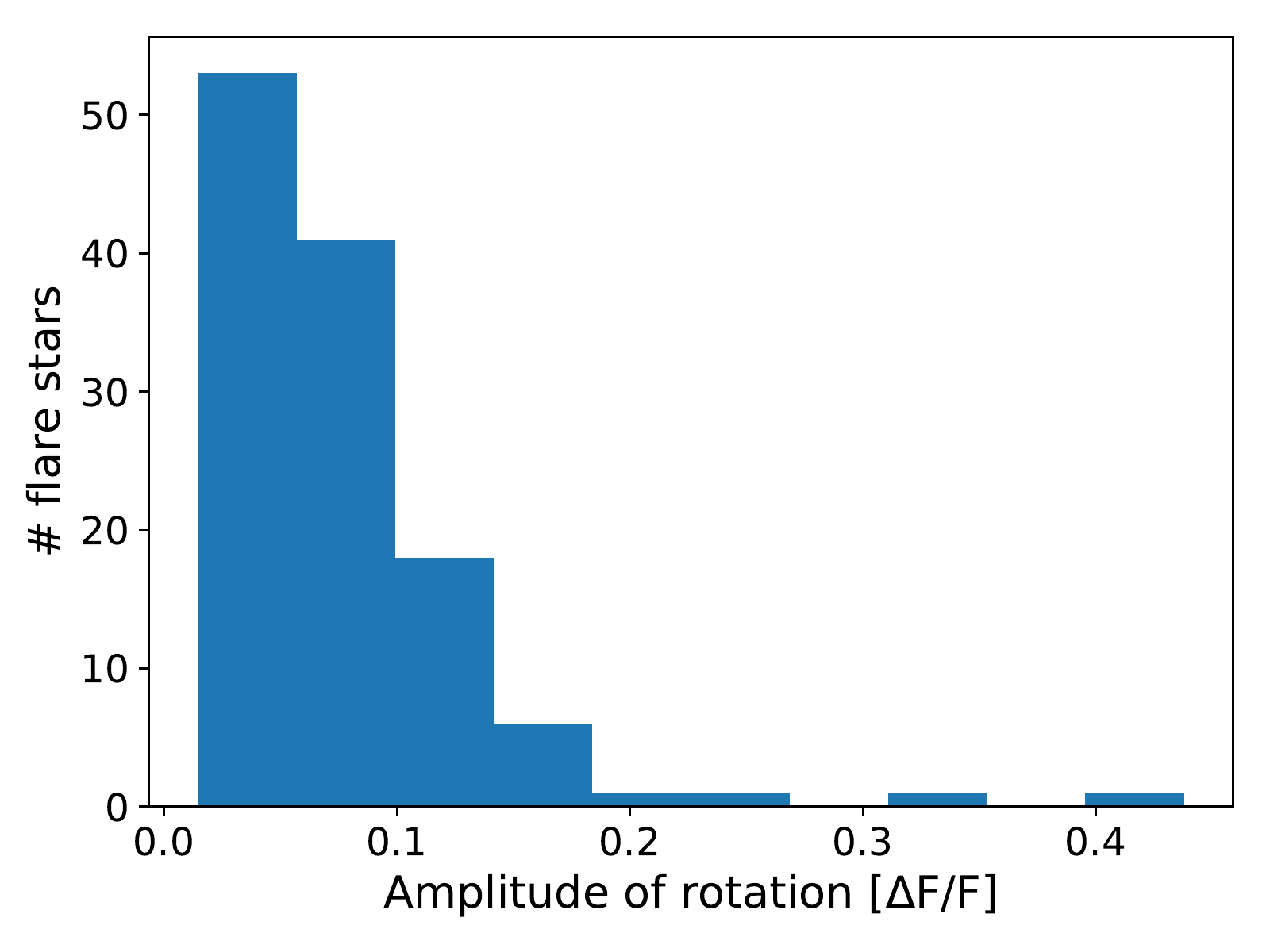}
		\label{fig:spots_overview_b}
	}
	\subfigure
	{
		\includegraphics[trim= 5 10 10 0, clip, width=2.25in]{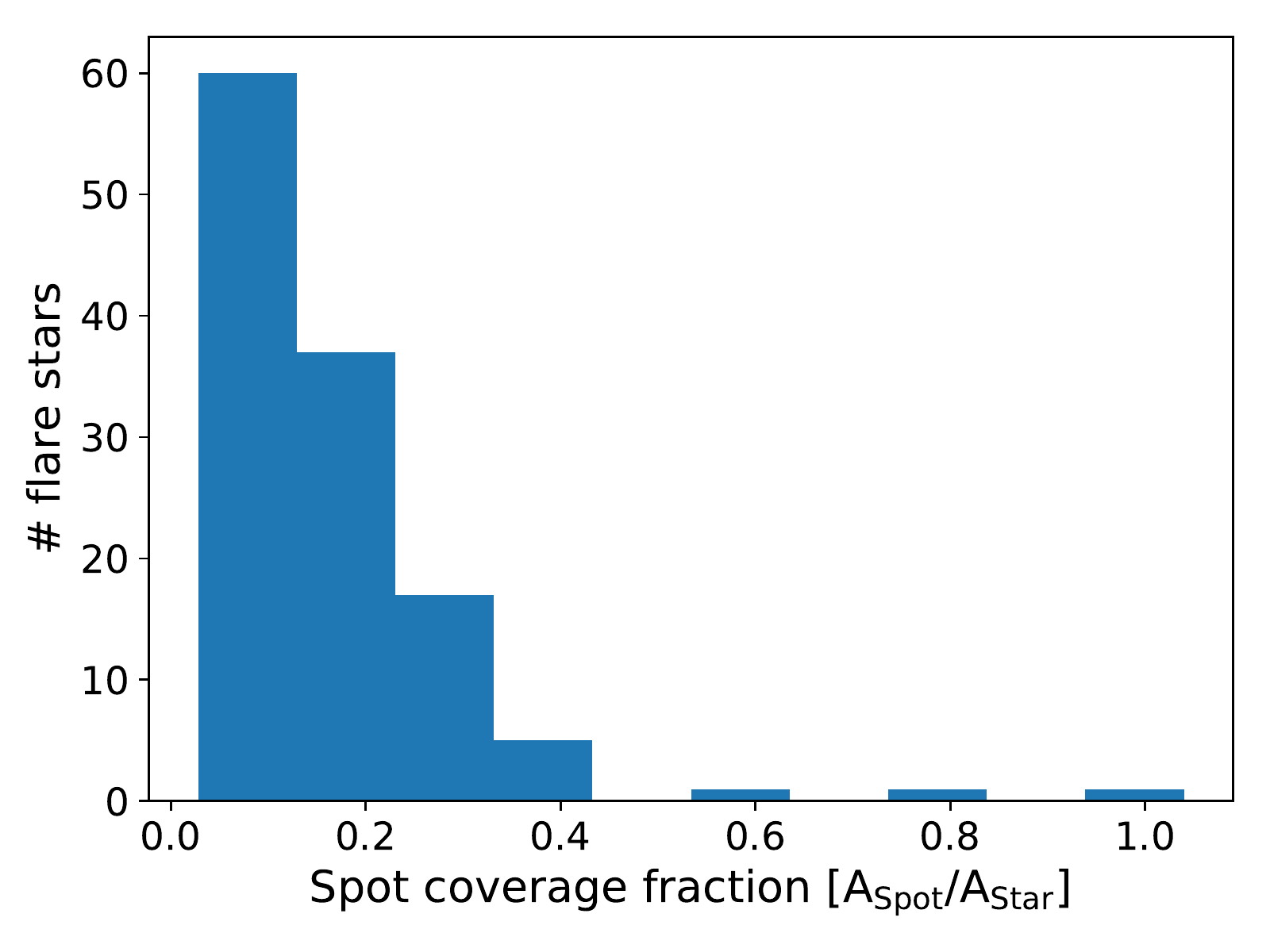}
		\label{fig:spots_overview_c}
	}
	\caption{Left panel: Histogram of the amplitudes of rotation detected by Evryscope, with a median amplitude and 1$\sigma$ spread in the distribution of amplitudes of $0.033\protect\substack{+0.026 \\ -0.014}$ $g^{\prime}$ magnitudes. Middle panel: Same as left panel, except in normalized flux units $\Delta$F/F, or the fraction of light blocked by spots, with a median amplitude and 1$\sigma$ spread in the distribution of normalized flux amplitudes of $0.06\protect\substack{+0.05 \\ -0.03}$. Right panel: Histogram of the distribution in hemispherical starspot coverage fraction, with a median spot coverage fraction and 1$\sigma$ spread in the distribution of spot coverage fraction of $0.13\protect\substack{0.12 \\ -0.06}$.}
	\label{fig:results_spot_ampl_overview}
\end{figure*}

We measure a distribution of sinusoidal amplitudes ranging from 0.008 to 0.216 $g^{\prime}$ magnitudes, with a median amplitude and 1$\sigma$ spread in the distribution of amplitudes of 0.033$\substack{+0.026 \\ -0.014}$ $g^{\prime}$ magnitudes, as shown in the left panel of Figure \ref{fig:results_spot_ampl_overview}. We convert amplitude of rotation in $g^{\prime}$ magnitudes to the normalized peak-to-trough flux amplitude $\Delta$F/F, which may be understood as the fraction of starlight blocked by spots ($\Delta$F/F is mathematically equivalent to fractional-flux). The median flux amplitude and 1$\sigma$ spread in the distribution of normalized flux amplitudes is 0.06$\substack{+0.05 \\ -0.03}$ as shown in the middle panel of Figure \ref{fig:results_spot_ampl_overview}.

The fraction of starlight blocked by spots $\Delta$F/F is not equivalent to the hemispherical starspot coverage fraction A$_{Spot}$/A$_{Star}$. This is because starspot area depends on the temperature of the star and the temperature of the starspots as given in Equation \ref{eq:spot_cover_fraction}. We estimate spot coverage fractions ranging from 0.03 up to nearly an entire stellar hemisphere; the median spot coverage fraction and 1$\sigma$ spread in the distribution of spot coverage fractions is 0.13$\substack{+0.12 \\ -0.06}$. We note that coverage fractions depend on the assumed spot temperature, stellar radius, and fraction of bolometric spot flux observed in $g^{\prime}$, which may each be in excess of 10\% error; we urge readers to exercise caution in the use of these values where precision better than 50\% is required.

\begin{figure}
	\centering
	{
		\includegraphics[trim= 10 3 39 28,clip, width=3.4in]{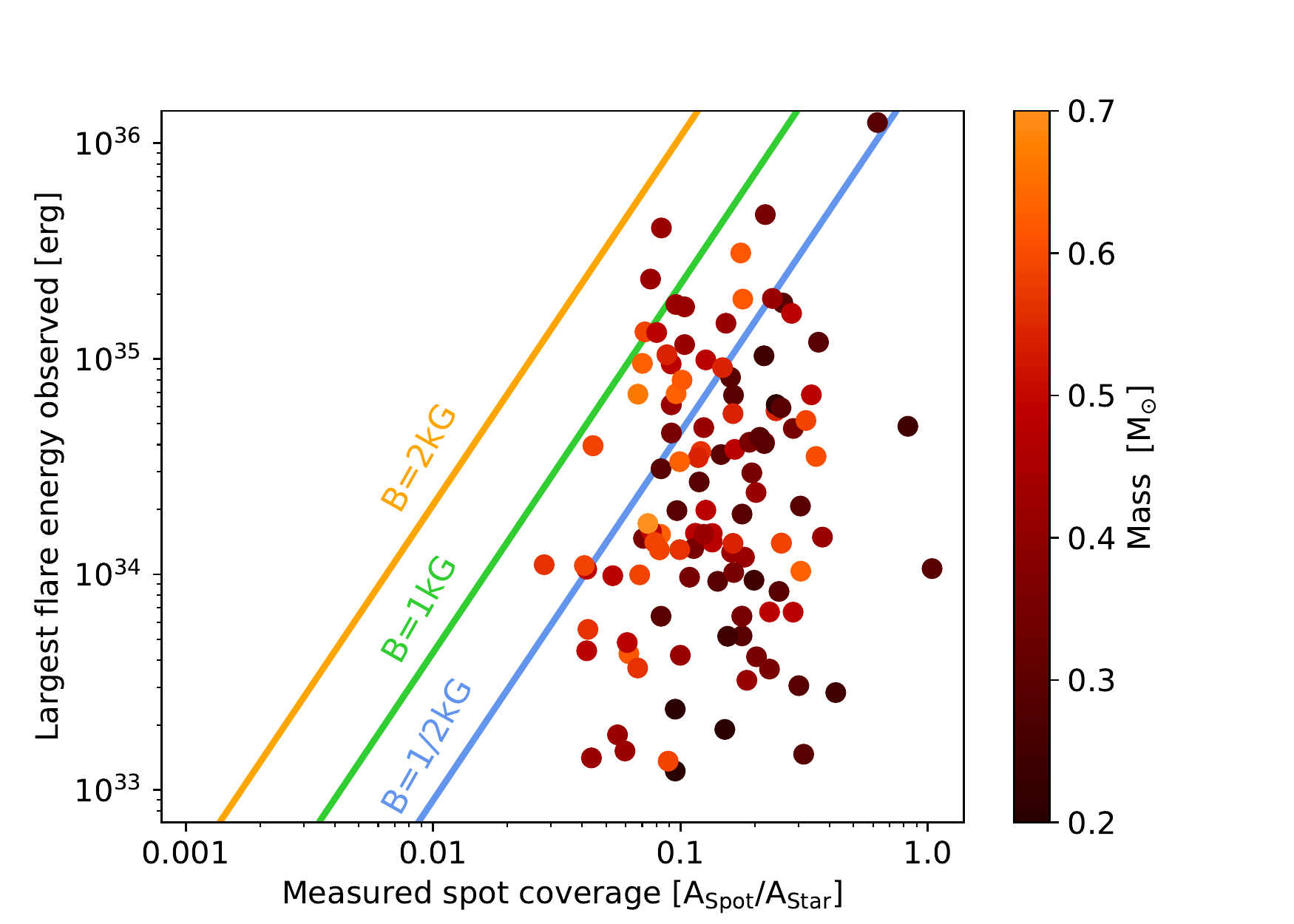}
	}
	\caption{Measured starspot coverage of each rotating star versus the maximum-observed flare energy from that star. Scaling relations for the minimum spot coverage needed to generate flares at the observed energies are overlaid for representative field strengths of 0.5 kG, 1 kG, and 2 kG. For each scaling relation for a particular field strength, the measured spot coverage should lie to the right of that line. We find most of our rotators lie to the right of the 0.5 kG field line, and all lie to the right of the 2 kG line, placing upper limits on the minimum field strength of our sample. We also color-code each data point representing a rotating flare star by its stellar mass, finding a gradient between early and mid M-dwarf stars in the plane of stellar mass and flare energy.}
	\label{fig:ampl_spots_vs_fieldlines}
\end{figure}

Energy stored in starspots may be released in the form of stellar flares. The area of the smallest spot that could have produced a flare of bolometric energy $E_\mathrm{flare}$ is given by \citet{Shibata2013,Notsu2019} as:
\begin{equation}
    E_\mathrm{flare}=\frac {B^2}{8\pi} A_\mathrm{Spot}^{3/2}
    \label{eq:notsu_spot_flare_scaling}
\end{equation}
$B$ is the surface magnetic field strength, and $A_\mathrm{Spot}$ is the smallest spot group area that could release a flare of energy $E_\mathrm{flare}$. We note that true spot sizes could be at least an order of magnitude larger than those given by this simplified model. We plot the largest flares we observed from each star as a function of the estimated starspot coverage of that star in Figure \ref{fig:ampl_spots_vs_fieldlines}. We then overlay lines of minimum starspot coverage capable of generating the maximum-observed flare energy from each star, for representative magnetic field strengths of 0.5 kG, 1 kG, and 2 kG as shown in Figure \ref{fig:ampl_spots_vs_fieldlines}. Because the true spot coverage ought to lie to the right of this line (i.e. greater spot coverage), we may constrain the minimum field strength $B$ associated with our starpots (in certain line-of-sight spot geometries, a smaller field could be several kG larger).

We find most stars in our sample lie to the right of the 0.5 kG field line, and all stars lie to the right of the 2 kG line. We therefore find a minimum magnetic field of 0.5 kG and a largest value for the minimum field strength of several kG, in broad agreement with previous measurements of the magnetic strengths of cool stars (\citet{Shulyak2017} and references therein). Interestingly, these field strengths are smaller than but comparable to those measured for rotating solar-type stars by \citet{Notsu2019}. We also note that Figure \ref{fig:ampl_spots_vs_fieldlines} shows a gradient in stellar mass across the plane of spot coverage versus maximum flare energy, implying late-type stars may sometimes have a smaller minimum field strength than earlier-type stars.

\subsection{Flaring and stellar rotation}\label{evr_rot}

\begin{figure*}
	\centering
	{
		\includegraphics[trim= 1 1 1 1,clip, width=6.9in]{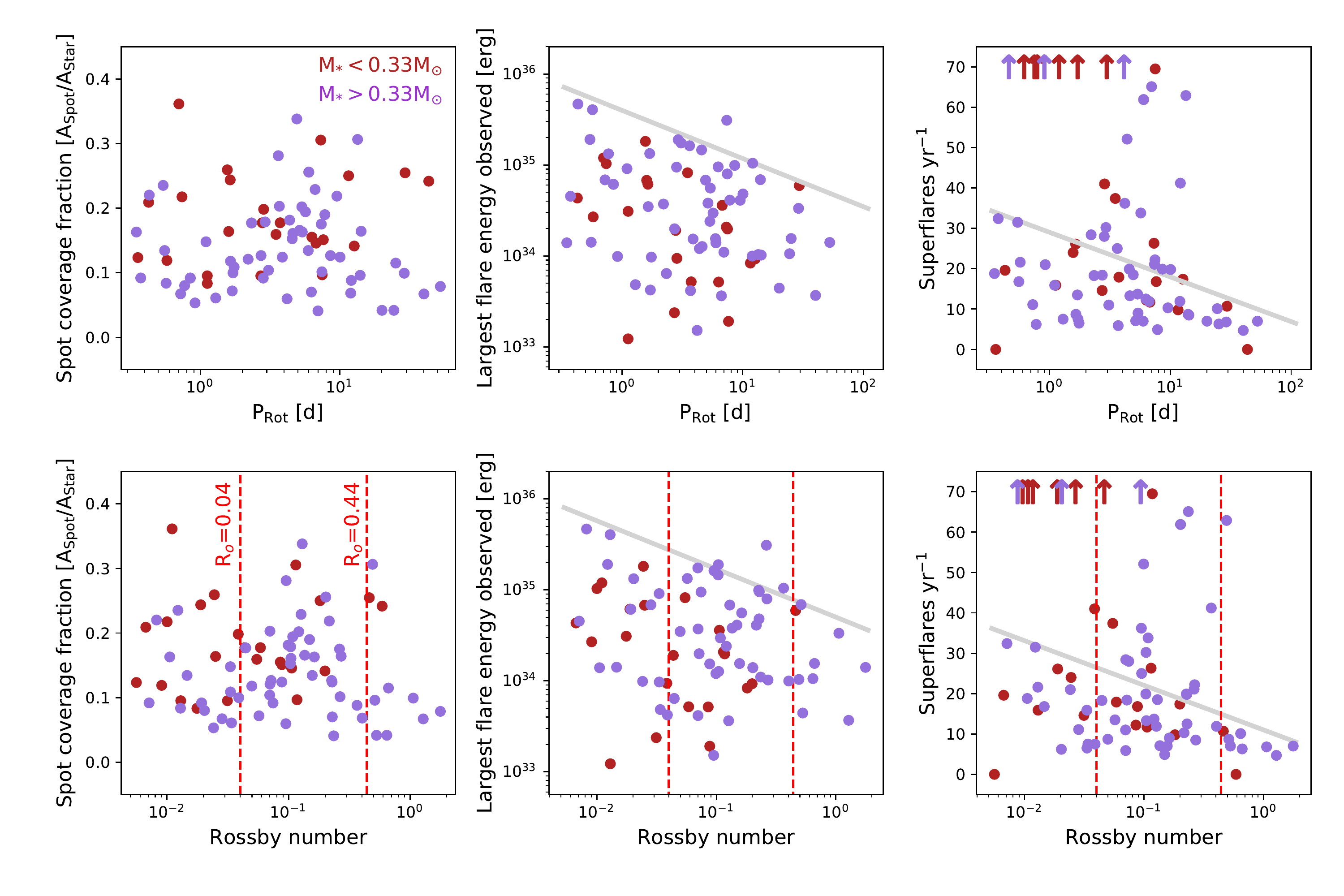}
	}
	\caption{Stellar activity observables as functions of stellar rotation and Rossby number. All points have periods confirmed in both TESS and Evryscope. Red points have stellar masses M$_{*}<$ 0.33 M$_{\odot}$, while purple points have stellar masses 0.33 $<$ M$_{*}<$ 0.7 M$_{\odot}$. Top panels: The starspot coverage fraction, largest observed flare energy from each star, and superflare rate versus rotation period. All three types of activity decrease at longer rotation periods, as described by Table \ref{table:activity_of_three_rot_popls_tab} and Table \ref{table:fast_inter_slow_rotators_tab}. To guide the eye, a grey line is overlaid on the decrease in stellar activity with period. The superflare rate changes significantly between periods of roughly 3 to 11 d. Bottom panels: The starspot coverage fraction, largest observed flare energy from each star, and superflare rate versus Rossby number. Vertical red dashed lines indicate the boundaries between the Rossby numbers of fast, intermediate, slow rotators. All three types of activity decrease at longer rotation periods, as described by Table \ref{table:activity_of_three_rot_popls_tab} and Table \ref{table:fast_inter_slow_rotators_tab}. However, the superflare rates of intermediate rotators show an apparent increase in flaring, if extremely-active stars (up arrows) are excluded. If real, this tentative evidence for changing surface magnetic field geometry during spin down may correlate with the increased activity of \citet{Mondrik2019}.}
	\label{fig:gridspec_flares}
\end{figure*}

\citet{Astudillo-Defru2017} and \citet{Newton2017} explore an increase in stellar activity as a function of rotation until the increase in activity shows saturation at periods shorter than $\sim$10 days. For those stars in our sample with recovered flares, we compare the amplitudes, energies and frequencies of their flares as a function of stellar rotation.

\subsubsection{Statistics of fast and slow rotators}\label{high_rates_MC}
We find an apparent increase in flare energy, amplitude, and superflare occurrence at short rotation periods, in general agreement with earlier results (e.g. \citet{Maehara2012,Paudel2018,Davenport2019}). However, some previous superflare surveys do not find any correlation of flare energy with rotation period, e.g. \citep{Maehara2012, gunter2019}. \citet{Maehara2012} suggest the maximum energy of a flare is thought to be dependent on the stored energy of a local active region, which does not necessarily depend on the stellar rotation. \citet{Notsu2019} report the \citet{Maehara2012} result is a result of giant contamination. More recently, \citet{Davenport2019} do find that flare strength decreases with increasing stellar rotation for all slowly-rotating cool stars. We note \citet{gunter2019} studied short-period rotators and \citet{Maehara2012} studied solar-type superflare stars instead of cool stars.

The relative difficulty in recovering long rotation periods means we may be sampling all activity levels at short periods and only the most common activity at long periods. This bias means that we must exercise caution in interpreting our results.

To correct for differences in stellar activity observables as functions of the rotation period, we group all recovered flare stars into $<$10 day (R$_{o}<$0.2) and $>$10 day (R$_{o}>$0.2) period bins of short-period and long-period rotators, respectively. We select these bins to directly compare our results to \citet{Astudillo-Defru2017} and \citet{Newton2017} who observed a break in rotation-activity power laws at this period. Looking ahead to Section \ref{flares_mass_rossby_plane}, we include the approximate Rossby number of a 10 d M-dwarf rotator because \citet{Astudillo-Defru2017} find a break in the power law describing M-dwarf activity versus period at 10 d but \citet{Newton2017} find the break at R$_o$=0.2. We hypothesize our short-period and long-period rotators are drawn from the same underlying distribution of superflare rates. Because we sample more short-period rotators than long-period rotators, we construct our random distribution of superflare rates based upon the observed distribution of short-period rotators.

We perform a Monte Carlo test of 10,000 trials with the goal of distinguishing if 79 short-period and 43 long-period rotators from the same simulated population can differ as much as our actual rotators do. In each trial, we simulate the same numbers of short-period rotators and long-period rotators as we actually observed, and test how often these simulated rotators differ as much as our observed rotators do by using the SciPy \citep{scipy2001} implementation of the two-sample Anderson-Darling (A-D) test \citep{Scholz1987}. 

All three stellar activity observables easily distinguish between our actual short-period and long-period rotators, with large A-D statistics and small p-values. This suggests they do not come from the same population. The MC trials support this interpretation: the A-D statistic and p-value of simulated rotators randomly drawn from the same underlying population do not distinguish between short and long-periods. Across 10,000 trials, the minimum p-values are 0.07, 0.06 and 0.04 and maximum A-D statistic values are 1.55, 1.71 and 2.33 for the superflare rate, maximum flare energy, and starspot coverage respectively. Since the simulated rotators cannot reproduce the difference in the activity of our actual rotators, we conclude the difference between our actual short-period and long-period rotators is unlikely to be due to sample bias. These results are shown in Table \ref{table:activity_of_three_rot_popls_tab}. We note running the same statistics excluding the 29 periods that do not correlate with TESS reduces the significance of the tests, although the activity-versus-period trends are still visible when only including periods confirmed in both surveys. See the top panel of Figure \ref{fig:gridspec_flares}.

\begin{table*}
\caption{Stellar activity of short period (P$_{Rot}<$10 d) vs. long period (P$_{Rot}>$10 d rotators}
\begin{tabular}{p{3.4cm} p{1.3cm} p{1.3cm} p{2.3cm} p{2.3cm} p{2.3cm} p{2.3cm}}
\hline
 & & & & & \\
Stellar activity observable & $p_\mathrm{obs}$ &  A-D$_\mathrm{obs}$  & Fraction trials \newline $p_\mathrm{sim}<p_\mathrm{obs}$ & $p_\mathrm{trials}$ \newline minimum value & Fraction trials \newline  A-D$_\mathrm{sim}>$A-D$_\mathrm{obs}$ & A-D$_\mathrm{trials}$ \newline maximum value\\
\hline
 &  &  &  &  &  & \\
Superflare rate & 3.2$\times$10$^{-5}$ & 13.12 & $<$10$^{-4}$ & 0.07 & $<$10$^{-4}$ & 1.55 \\
Largest flare energy & 1.0$\times$10$^{-5}$ & 17.52 & $<$10$^{-4}$ & 0.06 & $<$10$^{-4}$ & 1.71 \\
Spot coverage & 0.01 & 3.74 & $<$10$^{-4}$ & 0.04 & $<$10$^{-4}$ & 2.33 \\
 &  &  &  &  &  & \\
\hline
\end{tabular}
\label{table:activity_of_three_rot_popls_tab}
{\newline\newline \textbf{Notes.} We perform A-D tests on the stellar activity of our 79 short-period (P$_{Rot}<$10 d) and 43 long-period (P$_{Rot}>$10 d) rotators to distinguish if they arise from two distinct populations. We observe higher superflare rates, maximum flare energies, and starspot coverage from short-period rotators than long-period ones. While short-period and long-period rotators have distinct activity levels to significant p-values, we perform MC tests of 10K trials each to ensure our results are not entirely dependent on the larger number of short-period rotators. In each trial, we simulate the distribution of short-period rotators using acceptance-rejection sampling and draw the number of short-period and long period rotators we observed. We find that the fraction of the trials in which the A-D statistic and p-value of our simulated rotators more strongly distinguishes between short and long-periods than do the A-D statistic and p-value of our actual rotators is essentially zero. Across 10K trials, the minimum p-values of the simulated rotators are 0.07, 0.06 and 0.04 and the largest A-D statistic values are 1.55, 1.71 and 2.33 for the superflare rate, maximum flare energy, and starspot coverage respectively. The p-values of the observed rotators are more than an order of magnitude better (with the exception of spot coverage), and the A-D statistic values of the observed rotators are at least 60\% higher.} 
\end{table*}

\begin{table*}
\caption{Stellar activity of fast (R$_o<$0.04), intermediate (0.04$<$R$_o<$0.4), and slow (R$_o>$0.44) rotators}
\begin{tabular}{p{3.6cm} p{3.1cm} p{2.9cm} p{3.1cm} p{2.9cm}}
\hline
 &  &  &  & \\
Stellar activity observable & Fast vs. \newline intermediate $p_\mathrm{obs}$ &  Fast vs. \newline intermediate A-D$_\mathrm{obs}$ & Intermediate \newline vs. slow $p_\mathrm{obs}$ & Intermediate \newline vs. slow A-D$_\mathrm{obs}$ \\
\hline
 &  &  &  & \\
Superflare rate & 0.22 & 0.43 & 2.4$\times$10$^{-5}$ & 13.87 \\
Largest flare energy & 0.66 & -0.61 & 0.01 & 3.36 \\
Spot coverage & 0.28 & 0.21 & 0.003 & 5.05 \\
 &  &  &  & \\
\hline
\end{tabular}
\label{table:fast_inter_slow_rotators_tab}
{\newline\newline \textbf{Notes.} We perform A-D tests on the stellar activity observables of our 30 fast rotators (R$_o<$0.04), 59 intermediate-period rotators (0.04$<$R$_o<$0.4), and 33 slow rotators (R$_o>$0.44) to distinguish if they arise from distinct populations. We do not observe significant A-D statistic values or p-values between the stellar activity of our fast and intermediate rotators. We do observe a significant difference between the superflare rate and starspot coverage of the intermediate and slow rotators. The largest flare energies of the intermediate and slow rotators do not demonstrate significant differences, likely due to the small numbers of flare stars observed since the flares in Table \ref{table:activity_of_three_rot_popls_tab} do display a difference. We note that we do not conclusively confirm the higher activity of intermediate rotators detected in MEarth light curves by \citet{Mondrik2019}. We believe this to be a result of our sample size and urge future work with larger samples of cool stars.}
\end{table*}

\subsubsection{Quantifying rotation with the Rossby number}\label{flares_mass_rossby_plane}
In addition to the rotation period, stellar rotation is also quantified by the Rossby number: $R_o$=$P_\mathrm{Rot}$/$\tau_\mathrm{Conv}$, where $\mathrm{\tau}_\mathrm{Conv}$ is the convective turnover timescale in the star. R$_o$ gives the relative strength of Coriolis forces and inertial forces in the star (i.e. when the Rossby number is small, the star rotates quickly, and Coriolis forces have the greatest impact upon the surface magnetic field). Convective turnover time is calculated using Equation 11 of \citet{Wright2011}. This equation is valid in the mass range 0.09 $<$ M$_{Star}$/M$_{\odot}$ $<$ 1.36. Because the convection turnover time depends upon the stellar mass, inaccuracy in the determination of the mass used in calculating convection turnover timescale will be propagated to the Rossby number. In the cool star mass range, uncertainty in the stellar mass of 0.1M$_{\odot}$ can propagate to errors in the Rossby number of up to $\sim$0.15 dex.

We find 30 (24.6\%) of our flare stars to be fast rotators (R$_o<$0.04), 59 (48.4\%) to be intermediate-period rotators (0.04$<$R$_o<$0.4) undergoing rapid evolution to the topology of the surface magnetic field during spin-down, and 33 (27.0\%) to be slow rotators (R$_o>$0.44). We define fast, intermediate, and slow rotators this way to be consistent with the convention of \citet{Mondrik2019}. In Figure \ref{fig:massrot_vs_field}, we explore the stellar mass and Rossby number as functions of the spot coverage, maximum flare energy observed per star, and the superflare rate. We find our flare star sample explores the period-gap reported in earlier works (e.g. \citealt{Newton2018}).

\newpage
\subsubsection{Flare stars in the mass-Rossby plane}\label{flares_mass_rossby_plane}
We compare our rotators against rotators from other surveys. We plot low-mass and long-period rotators from the MEarth survey \citep{Newton2018}, and early M-dwarf to late K-dwarf rotators from the KELT survey \citet{Oelkers2018}. We convert the stellar effective temperatures from \citet{Oelkers2018} to stellar masses using the relations given in Table 5 in the Appendix of \citet{Kraus2007}. We find that Evryscope flare stars occupy a similar parameter-space in the mass-rotation plane as these surveys. However, our sample does not reach masses as low as some MEarth targets. What is unique about our sample compared to these MEarth and KELT targets is that our sample is selected on the basis of flaring, allowing us to probe changes in flaring in the mass-rotation plane.

We note the lack of fast rotators compared to intermediate rotators. We observe twice as many intermediate rotators as fast rotators. We check this lack is not a result of unexpected large errors in calculating $R_o$. Because our typical uncertainty in stellar mass is $\sim$0.1-0.2M$_{\odot}$ (i.e. a few spectral sub-types) can lead to errors in the Rossby number of up to 0.2-0.3 dex, our uncertainties are unlikely to account for the nearly order-of-magnitude difference necessary to move data-points between the the intermediate and fast rotator regime (visible as the bottom gray sequence below $R_o$=0.04 in Figure \ref{fig:massrot_vs_field}). We hypothesize that selecting rotators on the basis of a high flare rate is likely the cause of the high number of intermediate rotators. It is possible selection effects are present in Evryscope periodograms, suppressing the detection rates of fast rotation periods. Ruling out this possibility will require statistical analysis on a larger sample of Evryscope rotators that are not selected on the basis of flaring. 

Low-mass stars comprise the vast majority of fast rotators and therefore most of the fast rotators that have high superflare rates as shown in Figure \ref{fig:gridspec_flares} and Figure \ref{fig:massrot_vs_field}. In Figure \ref{fig:gridspec_flares}, we split our rotation-activity plots into low mass and high mass groups to determine if rotation-versus-activity changes across the fully-convective boundary. \citet{Mondrik2019}'s sample of flaring MEarth rotators are all M$_{*}<$ 0.33 M$_{\odot}$, motivating our choice of boundary.

\subsubsection{Inconclusive increased activity of intermediate rotators}\label{Ro_stat_tests}
We divide up all 122 rotating flare stars into fast, intermediate, and slow rotators and test if the stellar activity of the intermediate rotators is increased compared to the stellar activity of the fast and slow rotators. We perform 2-sample A-D tests as described in Section \ref{high_rates_MC} separately for the starspot coverage, maximum flare energy, and the superflare rate. We limit our hypothesis testing to three observables to avoid searching for random correlations. We choose observables that probe a broad range of stellar activity: a flare rate, a flare size, and the extensiveness of the active regions that emit flares.

For each observable, we test whether the fast and intermediate rotators come from the same population, and we test whether the intermediate and slow rotators come from the same population. We observe a general decrease in activity with decreasing rotation, in agreement with Table \ref{table:activity_of_three_rot_popls_tab} and earlier studies (e.g. \citet{Newton2017,Davenport2019,Ilin2019}). However, we do not statistically confirm the increased activity of intermediate rotators reported by \citet{Mondrik2019}. This is likely due to the small number of flare stars we observe; we urge more extensive studies of rotating flare stars be made. These results are displayed in Table \ref{table:fast_inter_slow_rotators_tab}.

We plot the stellar activity observables versus period and Rossby number in Figure \ref{fig:gridspec_flares} to verify the statistical results by visual inspection. While the statistical tests are performed on all 122 stars, we plot here only those stars with periods observed in both Evryscope and TESS. Although this cut removes some periods longer than the TESS observing window, it enables a simpler visual inspection of possible trends between the fast and intermediate rotator groups. We overlay grey lines indicating the trends in maximum activity versus rotation and search for excursions above these trend lines. There appear to be two groups of fast rotators, with one group showing lower superflare rates and the other group showing very high superflare rates. There is only one group of intermediate rotators, but this single group has a higher flare rate than the low activity group of fast rotators. It is possible the two groups of fast rotators evolve with age into the single group of intermediate rotators. 

\begin{figure}
	\centering
	\subfigure
	{
		\includegraphics[trim= 5 3 30 20, clip, width=3.4in]{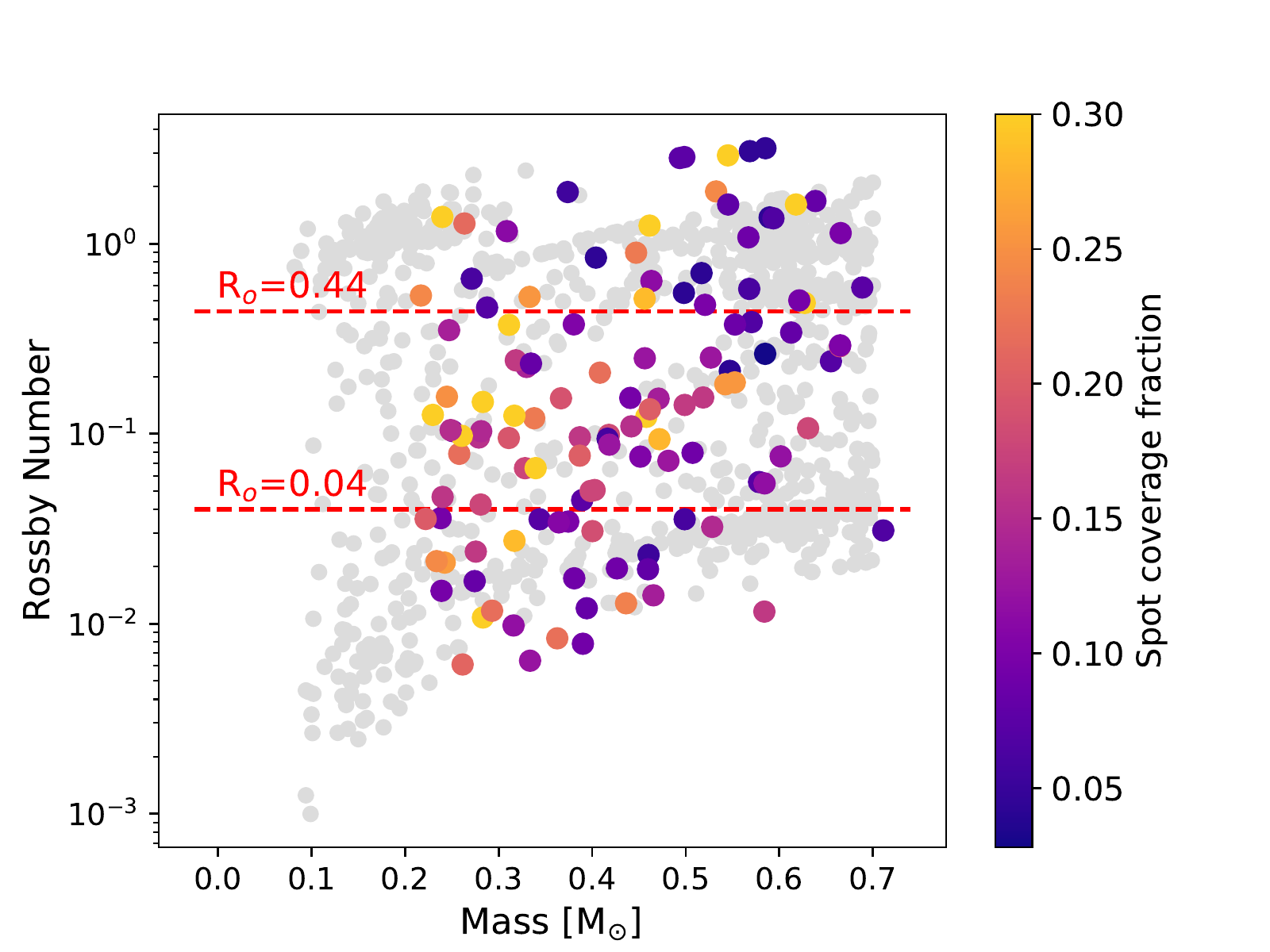}
		\label{fig:mass_vs_rot_f}
	}
	\subfigure
	{
		\includegraphics[trim= 5 3 30 20, clip, width=3.4in]{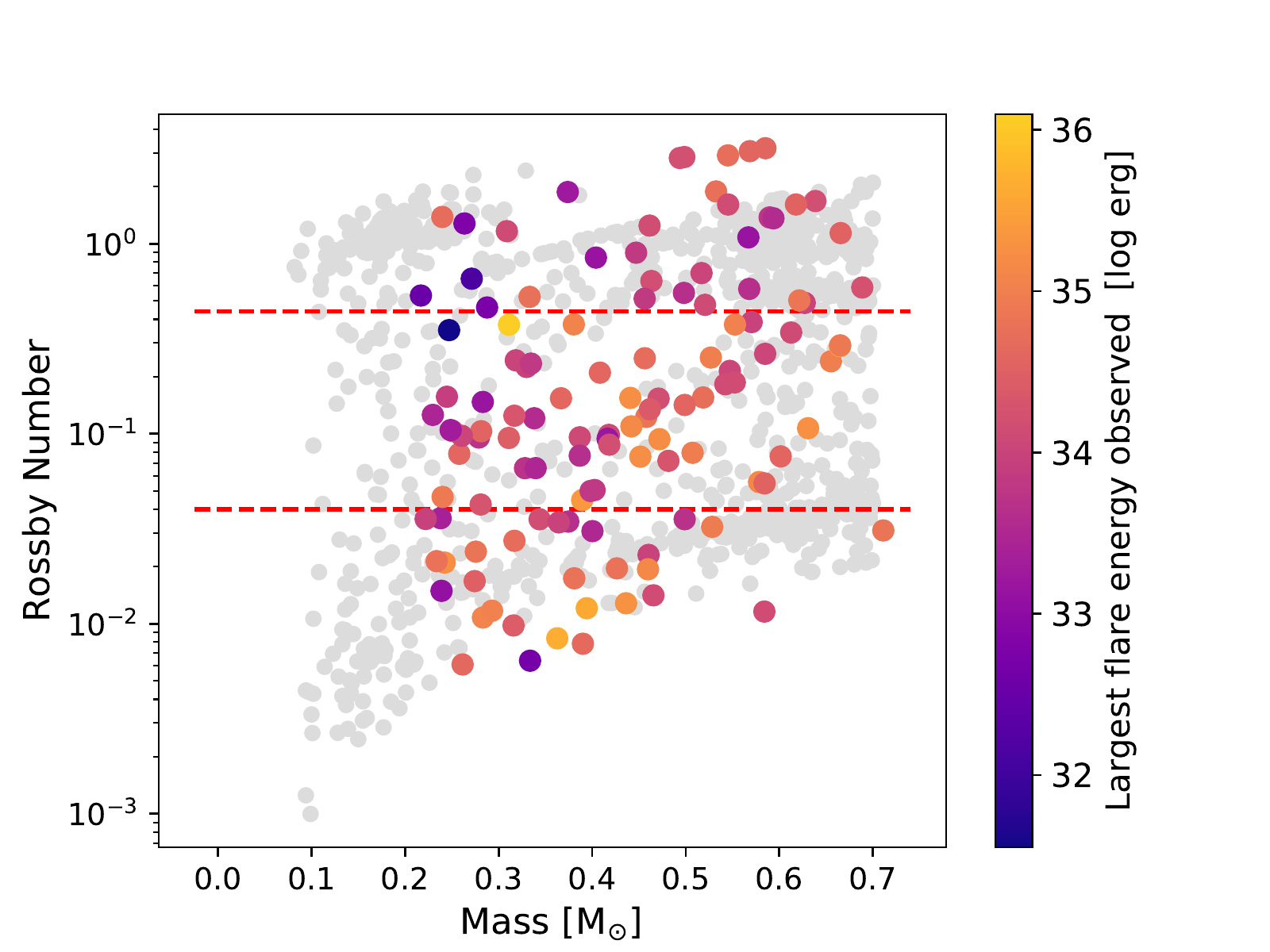}
		\label{fig:mass_vs_rot_c}
	}
	\subfigure
	{
		\includegraphics[trim= 5 3 30 20, clip, width=3.4in]{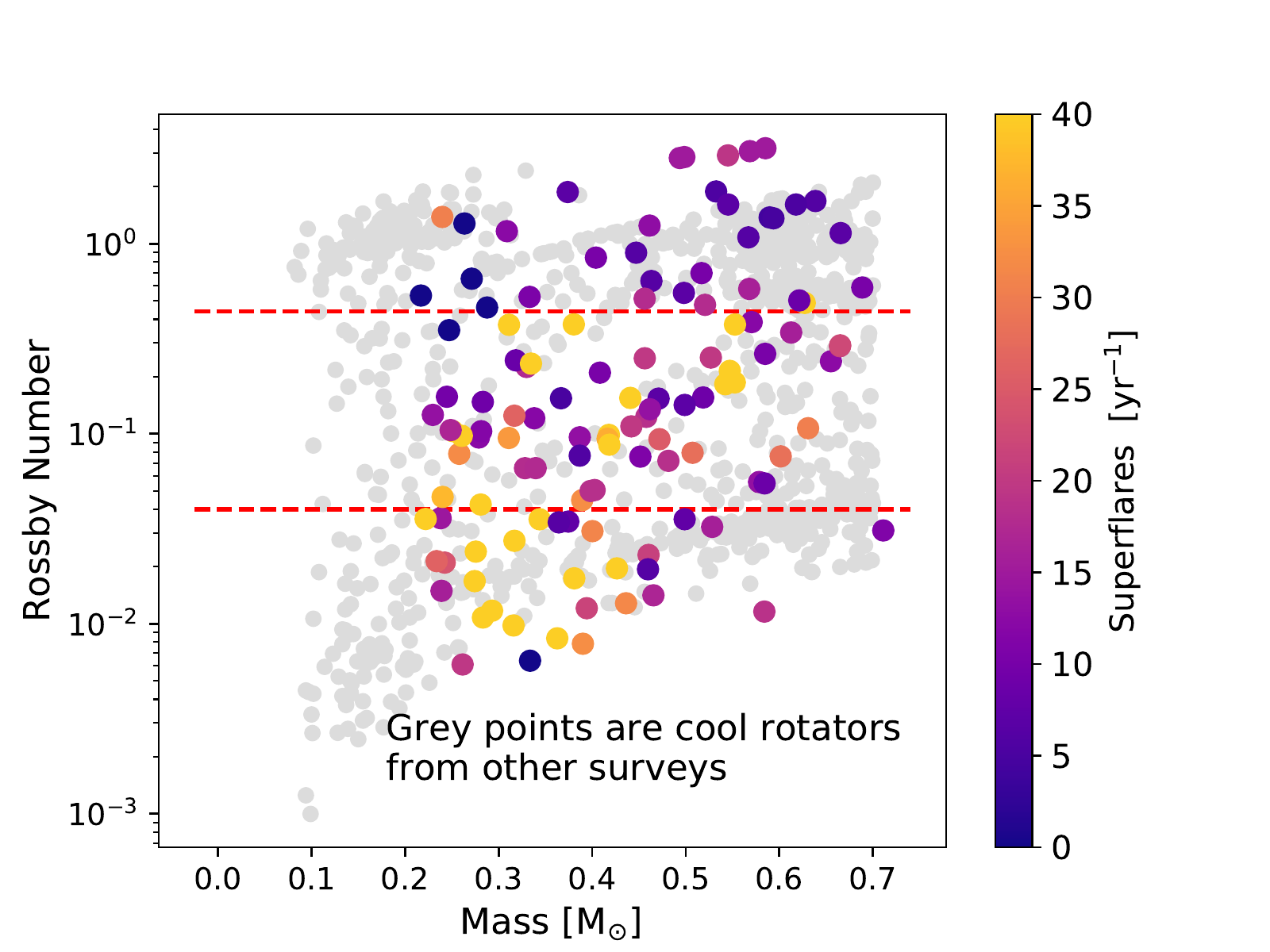}
		\label{fig:mass_vs_rot_d}
	}
	\caption{Evryscope flare stars in the mass-Rossby plane. Flare stars are scaled in color by (top panel to bottom) the spot coverage, maximum flare energy observed per star, and superflare rate, respectively. MEarth and KELT rotators not selected for flaring are plotted in grey for reference. Evryscope flare stars explore the spin down transition region from fast to slow rotation where \citet{Mondrik2019} report increased flaring.}
	\label{fig:massrot_vs_field}
\end{figure}

Our M$_{*}<$ 0.33 M$_{\odot}$ stars include both high and low activity groups of fast rotators and display the same patterns at longer periods as earlier-type stars. If a difference in mass between this work and \citet{Mondrik2019} explained their non-detection of the high-activity fast rotators, we would expect the high activity fast rotators to be earlier-type stars. However, Figure \ref{fig:mass_vs_rot_d} shows the high-activity fast rotators are mostly late-type stars. We urge further work with a larger sample of rotators and flare stars.

The spot coverage trend has high noise compared to the flare rate trend in Figure \ref{fig:gridspec_flares} and Figure \ref{fig:mass_vs_rot_f}. The maximum energies display a decrease with increasing period in Figure \ref{fig:gridspec_flares} and a diagonal gradient in the mass-Rossby plane of Figure \ref{fig:mass_vs_rot_c}.

\section{Summary and Conclusions}\label{discuss_conclude}
We observe 122 rotators in our sample of 284 late K and early-to-mid M flare stars, with periods ranging from 0.3487 to 104 days. We observe 30 fast rotators (R$_o<$0.04), 59 intermediate-period rotators (0.04$<$R$_o<$0.4) undergoing probable changes to the surface magnetic field, and 33 slow rotators (R$_o>$0.44). 

This sample of rotating flare stars was investigated as a subset of the ongoing Evryscope survey of all bright nearby stars; we selected these stars because they were observed in the first quarter of TESS observations and had 2-minute cadence light curves in the blue (Evryscope) and in the red (TESS), allowing future study of stellar activity in both bands. We find the sinusoidal amplitudes of rotation of cool stars often exceed 1\% variability, suggesting the combination of 28 d TESS observations and long-term, moderate-precision ground based observations may greatly increase the number and precision of rotation period measurements for nearby cool stars.

We fold the 2-minute cadence TESS light curve of each star to the Evryscope-detected period. We find the sinusoidal amplitude of rotation in the red TESS-bandpass is less than or equal to that in the blue Evryscope bandpass. We find this effect is strongest for the lowest mass stars in our sample and that the correlation with stellar mass is statistically significant.

Using the sinusoidal amplitude of rotation, we compute the minimum fraction of the stellar hemisphere covered by starspots. We measure a median spot coverage of 13\% of the stellar hemisphere. We predict the largest flares these spots could emit for several values of the stellar magnetic field strength and subsequently compare these large predicted flares against the largest flares we actually observed. We find stellar magnetic fields of at least 500 G are most-consistent with our observed flares and spots. The minimum field strength of the later-type cool stars exhibits a broader spread in values than the minimum field strength of the earlier-type cool stars.

We find our P$_{Rot}<$10 d (R$_{o}<$0.2) rotators demonstrate higher superflare rates, largest flare energies observed per star, and starspot coverage fractions than do P$_{Rot}>$10 d (R$_{o}>$0.2) rotators. Splitting up our rotators instead into fast (R$_o<$0.04), intermediate (0.04$<$R$_o<$0.4), and slow (R$_o>$0.44) rotators do not result in statistically significant increases from the fast to intermediate rotators, although a possible rise in the superflare rate of intermediate rotators is observed visually. Therefore, we do not conclusively confirm the increased activity of intermediate rotators seen in previous studies. Because our sample is specifically selected to only include flare stars from the 2 min cadence cool stars observed by TESS, the 2$\times$ increase in intermediate rotators we find over fast or slow rotators may itself be indicative of increased activity at these periods. However, this increase may be due to selection effects; we urge future work with larger samples of intermediate rotators be performed in Evryscope and TESS to confirm these apparent changes to starspot coverage during spin-down.

\section*{Acknowledgements}\label{acknowledge}
We would like to thank the anonymous referee who graciously gave their time to make this the best version of this work. 
WH thanks Derek Buzazi for a helpful discussion on well-known binary systems in the EvryFlare sample and Nicholas Mondrik for a helpful discussion on flaring MEarth rotators.
WH, HC, NL, JR, and AG acknowledge funding support by the National Science Foundation CAREER grant 1555175, and the Research Corporation Scialog grants 23782 and 23822. HC is supported by the National Science Foundation Graduate Research Fellowship under Grant No. DGE-1144081. OF and DdS acknowledge support by the Spanish Ministerio de Econom\'{\i}a y Competitividad (MINECO/FEDER, UE) under grants AYA2013-47447-C3-1-P, AYA2016-76012-C3-1-P, MDM-2014-0369 of ICCUB (Unidad de Excelencia `Mar\'{\i}a de Maeztu'). The Evryscope was constructed under National Science Foundation/ATI grant AST-1407589.
\par This paper includes data collected by the TESS mission. Funding for the TESS mission is provided by the NASA Explorer Program.
\par This work has made use of data from the European Space Agency (ESA) mission {\it Gaia} (\url{https://www.cosmos.esa.int/gaia}), processed by the {\it Gaia} Data Processing and Analysis Consortium (DPAC, \url{https://www.cosmos.esa.int/web/gaia/dpac/consortium}). Funding for the DPAC has been provided by national institutions, in particular the institutions participating in the {\it Gaia} Multilateral Agreement.
This research made use of Astropy,\footnote{http://www.astropy.org} a community-developed core Python package for Astronomy \citep{astropy:2013, astropy:2018}, and the NumPy, SciPy, and Matplotlib Python modules \citep{numpyscipy,Jones2001,matplotlib}.

{\it Facilities:} \facility{CTIO:Evryscope}, \facility{TESS}

\bibliographystyle{apj}
\bibliography{paper_references}

\end{document}